\documentclass[12pt,preprint]{aastex}
\input epsf.tex
\begin{document}
%% Prints a large "DRAFT" diagonally across each page
%% Does not show up in TeXview
% \typeout{Prints "DRAFT" on each page; does not show in TeXView}
% \special{!userdict begin /bop-hook{gsave 200 30 translate
% 65 rotate /Times-Roman findfont 216 scalefont setfont
% 0 0 moveto 0.95 setgray (DRAFT) show grestore}def end}
\title{Two Classes of Hot Jupiters}
\author{Brad M. S. Hansen\altaffilmark{1} \& Travis Barman\altaffilmark{2}
}
\altaffiltext{1}{Department of Physics \& Astronomy, and Institute of Geophysics \& Planetary Physics, University of California Los Angeles, Los Angeles, CA 90095, hansen@astro.ucla.edu}
\altaffiltext{2}{Hendricks Center for Planetary Studies, Lowell Observatory, 1400 West Mars Hill Road, Flagstaff, AZ 86001, barman@lowell.edu}

%\slugcomment{\it 
%}

\lefthead{Hansen \& Barman}
\righthead{Two Classes}

\begin{abstract}
We identify two classes of transiting planet, based on their equilibrium temperatures and
Safronov numbers. We examine various possible explanations for the dichotomy. It may reflect
the influence of planet or planetesimal scattering in determining when planetary migration
stops. Another possibility is that some planets lose more mass to evaporation than others.
If this evaporation process preferentially removes Helium from the planet, the consequent
reduction in the mean molecular weight may explain why some planets have anomalously large
radii.
\end{abstract}

\keywords{planetary systems: formation; binaries: eclipsing; scattering}

\section{Introduction}

Ever since the discovery of short orbital period, Jupiter mass extrasolar planets (Mayor \& Queloz 1995; Butler et al. 1997),
the so-called `Hot Jupiters', their nature and origin have been of great interest. This interest has
only intensified with the discovery of transiting planets (Charbonneau et al. 2000; Henry et al. 2000),
allowing the measurement of planetary radii and more accurate planetary masses.
 The identification of
  physical trends in the data may help to understand the physics of these objects. 
 Some such trends recently identified
include 
 trends between planet mass (Mazeh, Zucker \& Pont 2005) or
 gravity (Southworth, Wheatley \& Sams 2007) 
and orbital period for the known transiting planets. The existence of such correlations presumably
indicates something about the physical nature of these planets. 

In \S~\ref{Trends} we investigate these correlations further, in order to identify the true
physical variables that underlie them.  We introduce a new set of planet interior
models in \S~\ref{Models} for the purposes of comparing with the data.
The resulting comparison to the observed trends is performed in \S~\ref{Fits}.

\section{Planetary Trends}
\label{Trends}

In order to understand the origins of the correlation between planet gravity or mass and 
orbital period (Figure~\ref{2P}),
we wish to examine the correlation further using variables that are more closely tied to the likely
 physics of these objects. Figure~\ref{gT}
shows the planet gravity plotted against the equilibrium temperature
\begin{equation}
T_{eq} = T_{eff} \left( \frac{R_*}{2 a } \right)^{1/2}, \label{Teq}
\end{equation}
for the 17 known transiting planets with mass $<5 M_J$.
 This leads to an interesting difference with respect to Figure~\ref{2P},
in that the distribution shows a more bimodal character, with a hint of a gap at fixed $T_{eq}$ between low and high gravity
planets. To understand this further, we show in Figure~\ref{OT} the same data, but now with gravity replaced by the
Safronov number
\begin{equation}
\Theta = \frac{1}{2} \left( \frac{V_{esc}}{V_{orb}} \right)^2 =\frac{a}{R_p} \frac{M_p}{M_*},
\end{equation}
where $V_{esc}$ is the escape velocity from the surface of the planet and $V_{orb}$ is the orbital
velocity of the planet about its host star.
The division into two separate groups is now quite striking. There are two transiting planets that are
discrepant. The planet HD147506b is considerably more massive than the others and lies off the top of the plot
in Figure~\ref{OT}. 
 The isolated point at the lower left is the Neptune-mass companion
to GJ~436b, whose measured radius is much below (Gillon et al. 2007) that of a H/He mixture, and may legitimately considered a different kind of object from the others. The rest of the planets define
two separate classes, differing by almost a factor of two in their characteristic values of $\Theta$ at fixed
$T_{eq}$. We designate the planets with $\Theta \sim 0.07 \pm 0.01$ as Class~I and those with $\Theta \sim 0.04 \pm 0.01$ as Class~II.

Inspection of the properties of the two classes reveals that the Class~II objects orbit the hotter host stars in
general,
but at greater orbital separations than the Class~I stars with similar $T_{eq}$. This can be seen
in Figure~\ref{Ta}, which compares the two parameters that determine the amount of planetary illumination. 
 Class~II also contains
most of the planets who appear to have radii that are too large (e.g. Bodenheimer et al. 2001; Guillot \& Showman 2002) 
compared to basic H/He structure models\footnote{TrES-3 has the largest radius of the Class~I planets. However,
the slower cooling associated with the larger mass/heat capacity means that the observed radius agrees with
the model at ages $\sim 10^9$~years but not much beyond that. Thus, whether this planet can be considered anomalous
or not is somewhat ambiguous.}. Nevertheless, the differences in planetary radii are not enough to explain
the differences in 
 $\Theta$, because the radius anomalies are only $\sim 10\%$m whereas the characteristic value
of $\Theta$ differs by almost a factor of two between the two classes. The principal differences between
the two classes are based on mass. 
The planets of Class~II are, on average,  less massive than the corresponding planets in Class~I, and orbit
stars that are generally more massive. One way
of illustrating this is shown in Figure~\ref{MT}, in which we plot the planet mass versus $T_{eq}$. We see two
different linear trends for the two classes, with the Class~II planets being systematically less massive at
equivalent $T_{eq}$. Thus, not only is there a mass difference between the two classes, but there is a trend
of planet mass with irradiation within each class.

\subsection{Formalising the two classes} 

Setting aside then the Neptune-mass planet around GJ436b and the massive, eccentric planet
around HD~147506 as different beasts, we are left with 16 planets,
which we divide henceforth into two broad classes, based on the split in our $T_{eq}$--$\Theta$ diagram. The two
classes, with the relevant parameters are given in Table~\ref{Split}. The quantity $T_{eq}$ contains
only stellar and orbit parameters, so the difference between the two classes in $\Theta$ is clearly
 determined by planetary properties, encoded in the 
ratio $M_p/R_p$. This fact is the principal empirical result that we wish to try and understand.
The fact
 that the planets which have been identified as being
anomalously large (e.g. HD209458b; OGLE-TR-10) fall mostly into Class~II (although others, such as HD149026b, also fall in
this class) is not responsible for the difference in $\Theta$, but may offer a secondary indication as to the
physical origin of this bimodality.

 In subsequent sections we want to try and determine the physical origin of this
difference between Class~I and Class~II. To do so, first we introduce the models that we will use to compare to the data.

\section{Planet Models}
\label{Models}

Our planet evolutionary models are based on the Henyey code of Hansen (1996; 1999), originally used to
describe white dwarf evolution. Applying these models to the planetary evolution calculation is straightforward
once one substitutes the correct input physics. The most important addition is the
incorporation of the proper atmospheric boundary conditions for irradiated giant planets.
 For this we adopt the results of calculations 
using the PHOENIX code (Hauschildt \& Baron 1999). The atmosphere models are used to generate a grid of boundary conditions ranging from effective temperatures of 2000~K to 50~K and surface gravities from 300~$cm.s^{-2}$ to $3\times 10^4 cm.s^{-2}$. The grid
was recalculated for each system, using the stellar host and orbital separation parameters in Table~\ref{Split}.
The irradiation from the host star was included self-consistently in the radiative transfer in the manner described
in Barman, Hauschildt \& Allard (2001, 2005). The incident stellar flux was assumed to be distributed uniformly about the dayside
hemisphere of the planet.
Unless otherwise stated, the planet abundances were scaled to match the
published metallicities of the host stars and molecular abundances were
calculated assuming efficient rainout of dust (i.e. cloud free) as
described in
Barman et al.  (2005).  In most cases, rainout of dust resulted in
severely
reduced TiO and VO atmospheric number densities.
 Other important physical inputs
are the use of the Saumon, Chabrier \& Van Horn (1995) equation of state for H/He mixtures (already incorporated
during the white dwarf applications) and the calculation of a detailed Rosseland Mean opacity table using the
same opacities as used in the PHOENIX atmosphere models. At high densities, in the metallic phase,
 the H \& He electrical conductivities were also included.

Figure~\ref{Verify} shows an example of code verification, in which we reproduce the modelled evolution (dotted line) of the 
extrasolar planet HD209458b by Baraffe et al. (2003), using exactly the same boundary conditions (upper solid line). We also
include a second model (lower solid line) which uses slightly different boundary conditions for the same system, calculated
in the so-called `rainout' approximation, rather than the `cond' approximation used by Baraffe et al.
 The residual
differences in the cooling curves likely reflect small differences in the Rosseland mean opacity tables used in the
two cases. The good agreement shows that the theoretical models are consistent, although the gross discrepancy with the
observed radius indicates that whatever underlying physics was missing from the Baraffe \& Burrows models is
also missing from ours.

Using these models, we can repeat the exercises carried out by various other groups in the literature
 (e.g. Baraffe et al. 2005; Burrows et al. 2007) 
 comparing our evolutionary models to the various planetary cases, using the proper stellar illumination to
calculate the boundary conditions, in the rainout approximation. Performing the same calculations with our code yields similar results, so we will
not reproduce all the results here. However, we do show, in Figures~\ref{Panel1} and \ref{Panel2} the comparison
between our models for eight specific planetary systems, four from Class~I and 
four from Class~II. In all cases, we calculate the models assuming
that the planet has a cosmic composition of 73\% Hydrogen and 25\% Helium, using boundary conditions calculated
with the correct illumination (using the appropriate separation and host stellar type for each object). In each case
we include two curves. The first curve (dotted) indicates the radius calculated in the standard manner from the
Henyey code. The second curve (solid) is calculated to include the additional `transit radius effect' (Baraffe et al.
2003; Burrows et al. 2003), which takes into account that the surface of optical depth unity for limb-grazing stellar
photons (which is what the transit observations actually measure)
 lies higher in the atmosphere than the location of the surface in our model, which is taken to be
the location where the Rosseland mean optical depth $\tau_R=10^4$. Thus, we add to the model radius the
vertical extent of the atmosphere between these two locations, taken from the Phoenix atmosphere models.
The striking thing to note is that the models all match the Class~I observations in that they 
either intersect the observed radius or lie above them (if the planet has a rocky core, the radius will be
smaller than a coreless model of the same mass). The same is not true for the Class~II planets, where the very same models that fit the
Class~I planets often fail to match the observed radii. This is the well-known anomaly that was first identified with
the detection of the first transiting planet, HD~209458b (Bodenheimer et al. 2001; Guillot \& Showman 2002). 
The fact that this phenomenon appears to be correlated with the difference between the two classes hints at 
a common physical origin for the anomalous radii and the difference in Safronov numbers.

\section{Possible Explanations}
\label{Fits}

We have identified a bimodality in the value of the characteristic Safronov number $\Theta$ for the bulk of the 
known transiting planets. This suggests that the physics that underlies the split is associated somehow
with the planets ability to gravitionally scatter, capture or retain material. In this section we wish
to examine how this might come about within our current understanding of planet formation, migration
and evolution.

\subsection{Could it all be a selection effect?}

The cautious reader might wonder whether the fact that we find anomalously large planets around the 
hotter (and thus more massive stars) might be the result of some kind of systematic error in the 
estimation of the stellar radius. After all, the planetary radii in transit are measured relative to
the stellar radius. However, these results come from a variety of groups and so it would require the
error to be the same across several independent determinations i.e. it would require some fundamental
community-wide misunderstanding of the structure of these stars. Figure~\ref{RR} shows the ratio of
radii versus the ratio of masses, i.e. the quantities that are actually determined by transit and radial
velocity measurements.
 We see that there is a clear distinction between
the two classes in this diagram, suggesting that the split cannot be resolved by simply rescaling 
$M_*$ and $R_*$ for individual objects.
 Furthermore, the mass trend goes in the opposite
direction. If the stellar radii were being biased systematically high, and then so would the stellar
mass. Since planetary masses are also measured relative to the stellar mass, one would also expect them
to be biased high. This is the opposite of the observed trend.

\subsection{Energy Redistribution}

One of the parameters we have used to classify the stars into two groups is the so-called `equilibrium
temperature $T_{eq}$. This is important because, for giant planets so close to their host stars, the planetary structure (most importantly, the global entropy)
is regulated by the irradiation it receives.
 To calculate this number we must make an assumption about how the incoming energy is
redistributed across the surface of the planet. The calculation presented in equation~(\ref{Teq}) assumes
that there is little redistribution over the surface of the planet, i.e. the area that is re-emitting with
average temperature $T_{eq}$ is $2 \pi R_p^2$. If the energy is redistributed efficiently across the
surface of the planet, that area would be $4 \pi R_p^2$ and the numerical value of  $T_{eq}$ would be
reduced by a factor $\sim 0.84$. Rescaling both classes by the same amount would obviously have no effect,
but if one class of planet had efficient redistribution and the other did not, it could move the groups
close together. There might even be some precedent, in that recent measurements of phase curves for
a few extrasolar planets suggest different degrees of redistribution on planetary surfaces. Knutson
et al. (2007) find that redistribution is quite efficient for the transiting planet HD189733b, a member of
Class~I. On the
other hand, the measurement of a phase curve for the non-transiting planet $\upsilon$~Andromedae (Harrington et al. 2006) suggests that redistribution is weak in the atmosphere of this planet. Although this system is not
transiting and so not formally within our classification, the host star in this system is of spectral type F8 and thus
plausibly a member of  Class~II\footnote{Since the inclination angle is unknown, we can place a lower limit on
$\Theta$ for $\upsilon$~And~b by taking the largest plausible radius, yielding $\Theta > 0.045$ for $R_p=1.5 R_J$.
An upper limit may be obtained by taking a small radius and using the inclination constraint from the observed
lightcurve ($i > 30^{\circ}$). This yields $\Theta < 0.134$ for $R_p=1 R_J$. Thus, formally we cannot rule
out membership in either Class.}. Furthermore, the large amplitude of the secondary eclipse in HD149026b (Harrington et al. 2007), a member of Class~II, suggests little redistribution.

 However, scaling the $T_{eq}$ of one class would only result in horizontal motion of one of the groups in Figure~\ref{OT}
and would not explain the difference in $\Theta$ values. Even scaling  $T_{eq}$ for Class~II
down by  $\sim 0.84$ does not bring the two groups into alignment in Figure~\ref{MT}. Thus, differences in
how the energy is redistributed cannot completely explain the difference in the two classes. Finally, the
observations suggest that strong redistribution occurs in objects like
HD~189733b, a Class~I system, so that it is Class~I that should have  $T_{eq}$ scaled down, which would only increase the
separation between the two classes.

\subsection{Evaporation}

The detection of excess Ly~$\alpha$ absorption during primary transit of HD209458b (Vidal-Madjar et al. 2003) 
indicates that Hot Jupiter planets indeed lose some level of mass. However, the inferred lower limit on
the mass loss rate is $\sim 10^{10} g.s^{-1}$, which is not enough to change the mass significantly. Before
we discuss the details of possible evaporation scenarios, we first wish to explore whether evaporation can
simultaneously explain the smaller masses and the larger radii of the Class~II planets. The planetary radius is not 
particularly sensitive to mass, but what sensitivity there is results in an inverse relationship i.e. lower
mass planets have larger radii than more massive planets under similar conditions. In particular, larger planets
cool less rapidly, so one might imagine that the radius, which is essentially a measure of the planet's global
entropy (e.g. Burrows \& Liebert 1993), would be larger if the planet spent a significant fraction of it's
lifetime with a higher mass than the present day mass.
 Can the lower masses of
the Class~II planets also explain their larger radii?

Figure~\ref{Evap} shows the radius evolution (solid line) of a planet that has it's mass reduced at a constant rate
from $1.1 M_J$ to $0.62 M_J$ over the course of 3~Gyr, while being subjected to illumination appropriate
to HD~209458b. This mass reduction is of the order required to move an object from Class~I 
to Class~II.
The dotted line corresponds to a planet that starts off at $0.62 M_J$ and the dashed
line corresponds to a planet where the mass loss is maintained at the same rate until the
planet is completely eroded. We see that the evaporation leads to
a slightly larger radius at intermediate ages, but that it only increases the planet radius by a few percent.
It seems as though simple evaporation alone cannot explain the anomalous radii as a holdover from a 
prior, more massive and hotter state.

\subsection{Tidal Heating}

Another scenario that has been suggested for stopping the inward migration of planets is that
the planet eventually overflows its Roche lobe (Trilling et al. 1998), possibly aided by tidal heating (Gu, Lin \& Bodenheimer 2003).
Could some process associated with this phenomenon give rise to the difference between the two
classes? Indeed, Ford \& Rasio (2006) note that the distribution of planetary mass with orbital
period appears to be confined to lie outside a locus defined by twice the Roche limit for each
system. This would arise if the final planetary configuration resulted from circularisation of
an initially eccentric orbit, whose periastron was located at the Roche limit. Figure~\ref{MMA}
shows the distribution of transiting planet mass ratios against semi-major axis. We see that,
although the Ford \& Rasio locus approximately delimits the observed sample, the trend is
better fit by lines of constant Safronov number, as would be expected based on Figure~\ref{OT}.
Certainly, the Roche limit does not seem to offer any way to explain the difference between the
two classes of planet.

Could the inflation by tidal dissipation have an additional influence?
In Figure~\ref{Tide} we show the tidal luminosity required to inflate each planet to fill
it's Roche Lobe at it's current position. This amounts to the assumption that the planet evolves
outwards due to mass loss through the Lagrangian L1 point, stopping at its present location when
the planetary radius detaches from the Roche lobe. We see that there is some correlation between 
$L_{tide}$ and $\Theta$, as to be expected since Class~I contains more massive planets on average,
which in turn require more heating to inflate to a given radius. However, there is significant
overlap between the two classes in terms of $L_{tide}$, suggesting that this too is not the defining
difference in the two classes.

\subsection{Accretion of Planetesimals}
\label{Accrete}

It is perhaps illuminating that the split into two classes occurs when plotted in terms of the
quantity $\Theta$, the Safronov number. This quantity is directly linked to the efficiency with
which a planet gravitationally scatters other bodies (e.g. Safronov 1972). The most commonly
accepted mechanism for planetary migration is loss of angular momentum due to torques in a 
gas disk (Lin, Bodenheimer \& Richardson 1996). In the absence of such a gas disk, planets may
still move inwards by either scattering other planets (Ford \& Rasio 1996) or planetesimals
(Murray et al. 1998). In particular, the planetesimal migration mechanism provides a natural
stopping mechanism in that the planetesimals eventually collide with the planet rather than
get ejected; a mechanism regulated by the Safronov number.

Following Tremaine (1993), we can estimate the stopping criterion as follows. Scattering of
planetesimals by a planet is a diffusive process, in which numerical integrations yield an
energy diffusion coefficient per orbit (e.g. Duncan et al. 1987)
\begin{equation}
 D_E \sim  \frac{10 Au}{a} \frac{M_p}{M_*}.
\end{equation}
Consequently, the random walk in energy requires that a planetesimal undergo, on average,
$N \sim (1/a D_E)^2$ periastron passages before being ejected. The probability that the planetesimal
strikes the planet on any given periastron passage is $ p \sim (R_p/a)^2/ \sin i$,
where $i$ is the inclination angle of the planetesimal orbit. Setting $N p \sim 1$ yields a
value for the critical Safronov number
\begin{equation}
\Theta_{crit} \sim 0.26 \left( \frac{ \sin i}{0.1} \right)^{-3/4}.
\end{equation}
Thus, $\Theta \sim 0.07$ requires a planetesimal disk with an opening angle $\sim 30^{\circ}$.
Smaller $\Theta$ require even larger opening angles, with a maximum $\theta \sim 0.046$, close
to the characteristic value of Class~II.

This scenario offers several ways to get two classes of planet orbit. Perhaps the simplest possibility
is that one class migrates via gas torques and another by scattering. If all planets migrate by scattering,
then the separation in $\Theta$ requires that the population of planetesimals scattered by Class~II planets
be dynamically `hotter' than those for Class~I planets.
 Alternatively, a two planet system
scattering planetesimals can result in an effective repulsion between the two planets (this is thought
to explain the outward migration of Uranus and Neptune in our own solar system e.g. Fernandez \& Ip 1984).
Perhaps the Class~II planets have a second body in the system, that results in them being pushed in further.
 Another possibility is that the difference results from the Class~I 
planets accreting a larger fraction of their mass in heavy elements as the migration slowed due
to their inability to eject planetesimals from deep within the potential well. Finally, we note that
the above diffusion coefficient comes from calculations performed for Oort~cloud applications, where
planetesimal accretion is a much smaller effect than it is likely to be in this case (Hansen 2000). 
The likelihood of these various scenarios can be better estimated once this calculation has been revisited.

If the Class~I planets accrete a significant amount of rocky material, that may offer an explanation for why some
radii, but not all, are anomalous. Most models for the anomalous radius of HD209458b and other
extrasolar planets postulate an extra source of energy of some kind (e.g. Lin, Bodenheimer \& Mardling 2001; Showman \& Guillot 2002; Winn \& Holman 2005) or a way to retard
the cooling more than normal (Chabrier \& Baraffe 2007). If we postulate that such an energy source
is present in {\em all} planets, we can explain the now smaller radii of the Class~I planets by virtue
of the fact that they have significantly supersolar metallicities, which means that their mean molecular
weights are larger and their radii correspondingly smaller\footnote{Burrows et al. 2007 have recently suggested
that increased metallicity may make planets {\em larger}, but they only increased the atmospheric opacity, without
increasing the mean molecular weight in the envelope. Inclusion of this latter effect, necessary for a self-consistent model, acts to make the radius
of the planet smaller, not larger, as one increases the metallicity.}. This is qualitatively similar to the proposal
made by Guillot et al. (2006), in which they postulate that a fixed fraction (0.5\%) of the irradiance energy is
somehow 
`recycled' as internal energy, albeit by an unknown mechanism. 

To examine this in more detail, we recalculate our models including an additional heating term. The extra
energy is deposited throughout the planet, proportional to local mass fraction. To illustrate the resulting effect, 
we recalculate the evolution of HD209458b. In order to match the observed radius, 
 we need to postulate an
internal luminosity $\sim 2.3 \times 10^{26} \rm ergs.s^{-1}$, which amounts to $1.7 \times 10^{-3}$ of the 
absorbed stellar flux. We can do a similar calculation for each of the transiting planets. There are
now two adjustable parameters (once one assumes the mass and irradiation based on the observations) --
the amount of extra heating and the core mass. The first variable makes the planet more extended, while
the second makes the planet more compact. Thus, one can trade off the two against one another in order
to fit the observed radius. Figure~\ref{EC}
demonstrates this for four representative planets in our sample. Each curve represents the relationship
between the extra heating $\epsilon$ and the core mass (in earth masses), determined by matching the observed
radius of the planet. Here $\epsilon$ is defined as a fraction of the irradiance for each planet, in the 
spirit of Guillot et al. (2006).
 In some cases, $\epsilon \rightarrow 0$ at finite core mass. This occurs when the observed
planet radius is smaller than the expected radius for a coreless model with no extra heating. On the other
hand, some models also require a finite $\epsilon$ even at zero core mass. This occurs when the observed
radius is anomalously large.
All models in Figure~\ref{EC} are calculated in the asymptotic limit (the extra heating contribution sets
a floor in the entropy and hence radius below which the planet does not contract). Thus, TrES-2 has a
non-zero $\epsilon$ at fixed mass even though we did not classify it as anomalously large based on it's
current age limit of $>1$~Gyr. 

Examination of Figure~\ref{EC} allows us to assess the viability of this model for explaining the observed
difference in Safronov number/Mass. If we follow the Guillot et al. notion of a fixed irradiance fraction
$\epsilon$, we can read off the core mass required. For example, a value of $\epsilon \sim 3\pm 2 \times 10^{-3}$,
implies a core mass for TrES-2 of $33 \pm 13 M_{\oplus}$, which is only $\sim 10\%$ of the total planet mass.
Thus, in order for TrES-2 to have accreted almost 50\% of its mass in heavy elements (as required to move
it from Group~II to Group~I), the internal energy source would have to be much stronger than is usually
assumed. A similar argument can be made for other Class~I sources, such as TR-113, which requires
a core mass $98 \pm 18 M_{\oplus}$, which is $\sim 23\%$ of the total. In summary, the scenario in
which the Class~I planets accrete a large amount of planetesimals places even greater demands on the
unknown energy source than in the traditional case.

Could the accreted material supply it's own heat, via radioactive decay? Using the estimated energy
output and abundance of $^{40}\rm K$ for the Earth's core (Gessmann \& Wood 2002 and references therein),
one can estimate an energy input of
\begin{equation}
 L_{rad} \sim 2 \times 10^{21} ergs\, s^{-1} \left( \frac{X_K}{10^{-4}} \right) \left( \frac{M_{core}}{100 M_{\oplus}} \right),
\end{equation}
which is orders of magnitude too small to affect the structure of these planets.

Thus, while  it is
certainly possible to make a consistent model in which {\em all} planets have a mysterious energy
source but the Class~II objects have accreted a substantial amount of material to reduce their
radii, the results are unsatisfactory on two counts. One is that it still requires the existence
of a mysterious energy source, and the other is that the amount of mass accretion required for the
Class~I is 
prohibitive, requiring in turn a much stronger energy source than is invoked for the Class~II sources.

\subsection{Evaporation and Helium-poor Planets}

The model of \S~\ref{Accrete} invokes enhanced metallicity to reduce a planetary radius. The
radius is reduced whether the metals are concentrated in the core or spread throughout the envelope.
 This occurs
because the increase in metallicity increases the mean molecular weight of the envelope (or core) and
leads to a steeper radial pressure gradient and a more compact planet. It then follows that decreasing
the mean molecular weight can increase the planetary radius. Is it possible to increase the radii of
Class~II objects by reducing their mean molecular weight? Since the default mixture is one composed of
73\% Hydrogen and 25\% Helium, is it possible to enhance the Hydrogen fraction of the envelope and
thereby increase the radius? For objects in the partially degenerate regime, such as planets, the
radius scales $R \sim \mu^{-1}$, so that it should be possible to extract a
radius increase of approximately 20\% by replacing the Helium in the model planet with Hydrogen.
This is more than required to explain the radius anomalies of the Class~II planets (which are $\sim 10\%$).

As an example, Figure~\ref{He209} shows the effect of removing the Helium on the evolution of HD209458b.
We use a boundary condition for an atmosphere computed assuming a solar (2\%) metallicity but depleted
in Helium. We also reduce the Helium content of the envelope by the same fraction.
 We see that global He mass fractions $Y \sim 0.14 \pm 0.02$ are consistent with the 
observed radius. In order to fit the other anomalously large planets, HAT-P-1 requires a model with
 $Y_f < 0.14$ and for WASP-1, $Y_f<0.13$. However, not all Class~II planets require low Helium content.
For instance, TR-56, despite it's large radius, is well fit by the cosmic composition model, because
of it's larger transit radius increment (a consequence of the higher surface irradiation).
 
Thus, we have the possibility of solving both the anomalous radius problem and explaining the
difference between the two classes, if we invoke significant evaporation in which the material
that is lost carries a supercosmic abundance of Helium. To quantify this,
consider the following simple model. We start with a planet that has initial mass $M_0$, with
the correct cosmic proportion of Helium, $Y_0=0.25$. If we remove a certain fraction of Hydrogen
and Helium in proportion $\alpha = \Delta M_H/\Delta M_{He}$, we are left with a final mass
$M_f$ and final Helium mass fraction $Y_f$. The initial and final masses are related by
\begin{equation}
M_0 = \left( \frac{4}{3 - \alpha} \right) \left[ 1  - (1 + \alpha) Y_f \right] M_f.
\end{equation}

We know $M_f$ and determine $Y_f$ by fitting our evolutionary models. If we furthermore require that
$M_0 \sim 1.8 M_f$, in order to match the difference in mass between Class~I and Class~II, then we
can infer the require value of $\alpha$. For HD209458b, with $Y_f \sim 0.14$, this 
yields $\alpha \sim 1.58$, instead of the value $\alpha=3$ which would preserve cosmic composition.
This corresponds to a wind in which Helium is 39\% by mass or 14\% by number.

Figure~\ref{EvapHe} shows the effect of Helium-rich evaporation in an evolutionary model. We start
with a 1$M_J$ planet irradiated in the manner of HD209458b, and remove mass at a fixed rate until
the planet reaches $0.63 M_J$, after which mass loss is shut off. The mass removed is composed of
50\% Helium by mass, so that the global Helium mass fraction of the planet is reduced, reaching
$Y=0.094$ by the of the mass loss. We also show the effect of evaporation assuming a solar composition
wind. The planet actually expands as Y drops. This expansion allows us to match the observations
of HD~209458b.

The final point to note is that, if 
 the observed planet population were composed of systems in varying stages of evaporation, the
distribution in $\Theta$ would exhibit a continuous distribution, rather than the observed bimodality.
In order to match the observations, the postulated evaporation  must only last for a finite time,
shutting off when $\Theta \sim 0.04$.

\section{Discussion}

We have identified a curious bimodality in the distribution of transiting planet properties, in
that they (apart from two outliers which may be further distinguished on the basis of
other characteristics) seem to possess one of two distinct values of the Safronov number. We have
explored several reasons for the dichotomy, two of which seem the most promising.

The first possibility is that the Safronov number, which essentially measures the efficiency with
which a planet scatters other bodies, plays an important role in determining when a planet halts
its migration. This would emerge quite naturally in the case where some or all of the planets migrate
as a result of ejecting smaller bodies through repeated gravitational encounters. While this scenario
naturally explains the prominence of $\Theta$, it offers no convincing explanation for a 
secondary observation, namely that those transiting planets with anomalously large radii seem to
fall predominantly into Class~II, the class with smaller values of $\Theta$.

It is interesting that our split into Class~I and Class~II planets show some similarities with
the assertion of Burkert \& Ida (2007), that the distribution of planets in semi-major axis interior
to 1~Au is different for planets around the highest mass ($>1.2 M_{\odot}$) stars in the sample and that
the planets in those systems are also systematically of lower mass. These two features are also defining
characteristics of our Class~II planets, suggesting that the distinction carries over to greater separations
than is currently probed by the transiting planet sample. This might be interpreted as giving support to
the suggestion that multiple migration mechanisms are at work.

A second possibility is that the difference in $\Theta$ is a consequence of the fact that some
of the planets have lost a markedly larger fraction of their mass through some form of evaporation.
This scenario requires that many of the Class~II planets lose $\sim 50$--60\% of their initial mass (but
not much more). The fact that Class~II planets are found preferentially among those with hotter host
stars is consistent with this idea, in that planet evaporation models are quite sensitive to the amount of EUV radiation received (e.g. Lammer et al. 2007 and references therein).
The exact amount of evaporation is still difficult to calculate because it depends sensitively on the
amount of chromospheric emission from the host star, a quantity poorly known even for F and G stars.
This scenario also offers a way of explaining the anomalous radii, if we allow for the possibility
that the evaporation reduces the Helium abundance in the planet, i.e. the mass in the wind carries a 
larger than solar abundance of Helium. The consequent reduction in the mean molecular mass of the planet
results in a larger radius for a fixed entropy. The anomalously large planets can all be fit with Helium
mass abundances $Y \sim 0.14$. We note that this need not happen in all cases, since some of the
Class~II planets can be fit with normal abundance models.

How might such an evaporation occur? 
Inspection of the exosphere models of 
Yelle (2004) indicates that the evaporating wind from an irradiated planet consists of ionized Hydrogen
but neutral Helium (for the exosphere temperatures $\sim 10^4$K lie between the ionization temperatures
of the two atoms). Thus, if a planet possesses a magnetic field, the loss of hydrogen may be reduced
(although not completely cut off -- there is an atomic Hydrogen component to the wind loss as well) in
such a way that the composition of the mass lost may be enriched in Helium. An alternative
model is that Coronal Mass Ejections from the star may have a significant impact on the mass 
loss (Khodachenko et al. 2006),
again depending on the strength of the planetary magnetic field.
Whichever process is in operation must also 
switch off when it reaches a limiting state defined by our Class~II, since the observed
distribution is not continuous.
Thus, we expect a somewhat
different evaporation history than that outlined in Baraffe et al. (2005),in which planets evaporate 
down to Neptune masses.

Where do Jupiter and Saturn fit into this picture? Figure~\ref{OTJS} shows the evolution of $T_{eq}$
and $\Theta$ in the event that Jupiter or Saturn migrated closer to the Sun. We see that they could
easily match the position of either group, depending on their location. This also illustrates the
peculiarity of the gap between the two groups, as it would be perfectly reasonable to place Jupiter
within the gap, if it was located at distances $\sim 0.027$~Au. On the other hand, it is intriguing
that, if Jupiter and Saturn were placed independently at the same close distance from the Sun, the difference between their
values of $\Theta$ would be
reminiscent of the gap between Class~I and Class~II. Furthermore, it has long been claimed that
Saturn's atmosphere is depleted in Helium, although the exact value is still 
somewhat uncertain (Gautier et al. 2006). For many years this has been understood in terms of the
separation of Helium in the metallic Hydrogen core of the planet (Stevenson \& Salpeter 1977) in
order to explain Saturn's anomalous luminosity (e.g. Pollack et al 1977). Could similar physics be
responsible for the difference between Class~I and Class~II? Fortney \& Hubbard (2003) have
already investigated the role of Helium phase separation in irradiated extrasolar planets and have
concluded that the interior temperatures for the Hot Jupiters are likely too high for such processes
to be relevant. It therefore appears that the Helium paucity of Saturn's atmosphere and it's
likely membership in Class~II had it migrated inwards, may simply be a coincidence.

The last couple of years have seen a variety of measurements concerning
the properties of these transiting planets, many of which probe the state of the planet atmosphere directly.
The extreme difficulty of these measurements means that comparison to atmosphere models are essential and
thus the interpretations are necessarily linked with what one assumes for the model. We hope that some of
the above considerations will encourage modellers to take a very broad view of their input assumptions and
cover the widest possible range of parameter space, including both heavily metal enriched atmospheres and
atmospheres with strongly reduced Helium abundance.

\acknowledgements BH acknowledges support by NASA ATP contract NNG04GK53G and thanks Mike Jura, Andrea Ghez
Sara Seager and Scott Tremaine for comments and/or skeptical faces. 
TB acknowledges support from NASA Origins of Solar System
grant NNX07AG68-S02 and the Spitzer theoretical research
program.  We also acknowledge NASA's Advanced Supercomputing
Division for their generous allocation of time on the Columbia
supercomputer.

\newpage
\begin{deluxetable}{lcccccccc} 
\tablecolumns{8} 
\tablewidth{0pc} 
\tablecaption{The properties of the known transiting planets, based on the tabulation
of 
{\tt http://obswww.unige.ch/\~{}pont/TRANSITS.htm}, 
as of June 1~2007. Ages are taken from the compilation of Melo et al. (2006). Where no data
was available, we have assumed a lower limit of 1~Gyr for the age. \label{Split}} 
\tablehead{ 
\colhead{Planet Name} & \colhead{$\rm M_p$}   & \colhead{$\rm R_p$}    & \colhead{$\rm M_*$} & 
\colhead{$T_{eff}$} & \colhead{a} & \colhead{Age} & \colhead{T$_{eq}$} & \colhead{$\Theta$} \\
 & \colhead{ ($M_J$)} & \colhead{ ($R_J$)} & \colhead{ ($M_{\odot}$)} & \colhead{ (K)} & \colhead{ (AU)} &
(Gyr) & \colhead{(K)}  }
\startdata 
\cutinhead{ Class I} 
OGLE-TR-111 &  0.52 &  1.01  & 0.81 & 5044 &  0.047 & $>1.1$ & 1027 & 0.062 \\
OGLE-TR-113 &  1.35 &  1.09   & 0.77 & 4804 &  0.023 & $>0.7$ & 1345 & 0.076 \\   
HD189733b & 1.15 & 1.16 & 0.82  & 5050 &  0.031 & $>0.6$ & 1201 & 0.079 \\
TrES-1 &  0.76 & 1.08 & 0.89 & 5250 & 0.039 & 2.5$\pm$0.1 & 1151 & 0.065 \\ 
TrES-2 & 1.20 & 1.22 & 0.98 & 5850 & 0.037 & $>1$ & 1474 & 0.077 \\
XO-1  & 0.90 & 1.18 & 1.0 & 5750 & 0.049 & $>1$ & 1210 & 0.078 \\
WASP-2 & 0.88 & 1.04 & 0.79 & 5200 & 0.031 & $>1$ & 1292 & 0.069 \\
TrES-3 & 1.92 & 1.30 & 0.90 & 5720 & 0.023 & $>1$ &  1645 & 0.078 \\
\cutinhead{ Class II} 
OGLE-TR-10  & 0.61 & 1.22 & 1.10 & 6075 & 0.042 & $>1.1$ &  1535 & 0.040 \\
OGLE-TR-56  & 1.29 & 1.30 & 1.17 & 6119 & 0.023 & $3\pm 1$ & 2262 & 0.040 \\  
OGLE-TR-132 & 1.14 & 1.18 & 1.26 & 6210 & 0.030 & $>1$ & 2007 & 0.048 \\
HD149026b & 0.33 & 0.73 & 1.3 & 6147 & 0.042 & $2 \pm 0.8$ & 1743 & 0.031 \\
HD209458b  & 0.66 & 1.32 & 1.10 & 6117 & 0.047 & 4.5 & 1445 & 0.045 \\
HAT-P-1  & 0.53 & 1.36 & 1.12 & 5975 & 0.055 & $>1$ & 1318 & 0.040 \\
WASP-1 & 0.87 & 1.44 & 1.15 & 6110 & 0.038 & $>1$ &  1819 & 0.042 \\
XO-2 & 0.57 & 0.97 & 0.98 & 5340 & 0.037 & $>1$ & 1316 & 0.046 \\
\cutinhead{Unclassified}
HD147506 & 8.17 & 1.18 & 1.35 & 6290 & 0.069 & $>1$ & 1556 & 0.737 \\
GJ436 & 0.07 & 0.35 & 0.44 & 3200 & 0.028 & $>1$ & 612 & 0.027 \\
\enddata 
\end{deluxetable}

\newpage
\plotone{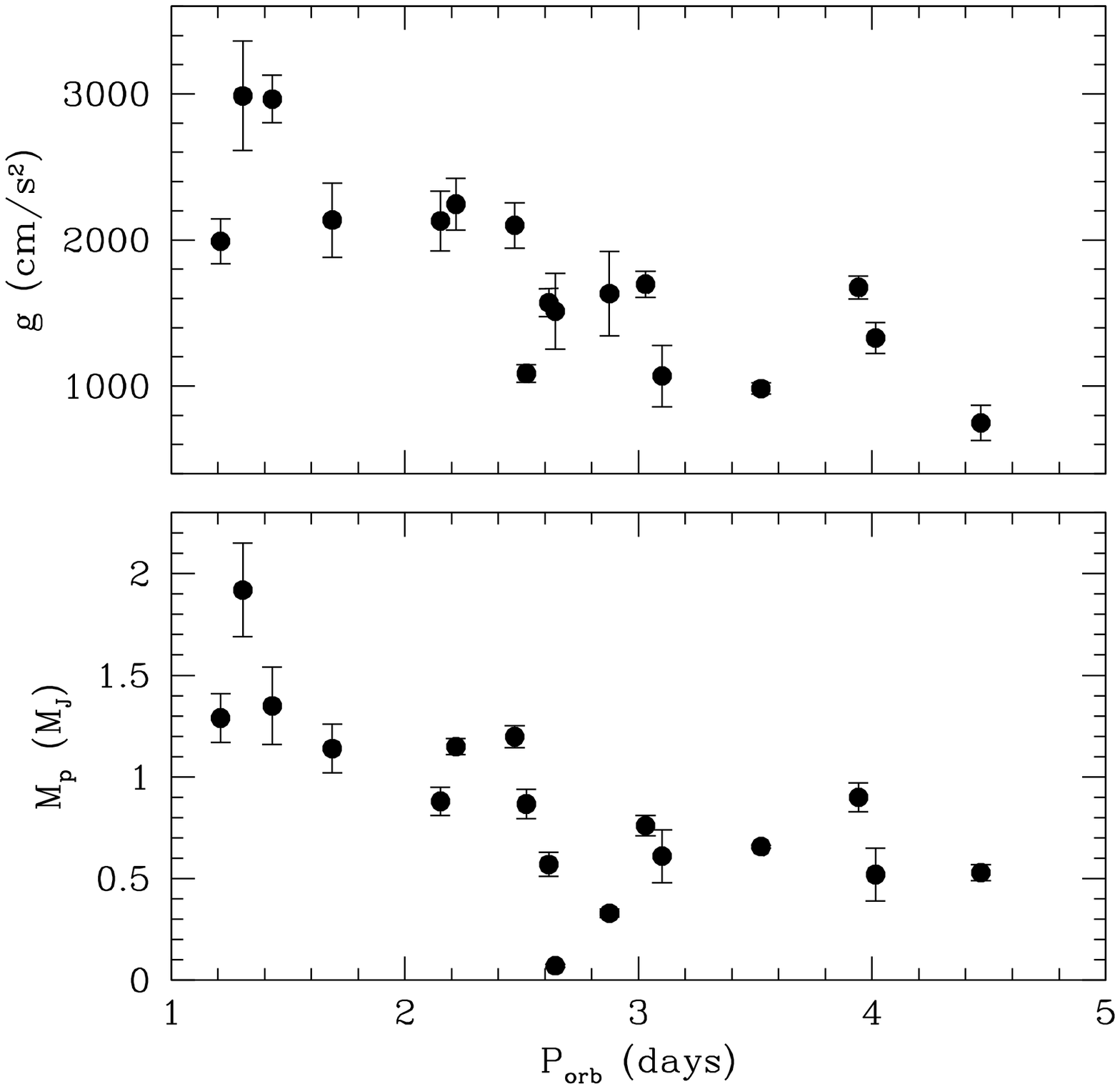}
\figcaption[f1.ps]{ The upper panel shows the planet gravity plotted against orbital period, for the known
 transiting planets. The lower panel shows the planet mass versus orbital period. In both cases a trend is
visible, with shorter period planets having higher masses \& gravities on average. \label{2P}}
\newpage
\plotone{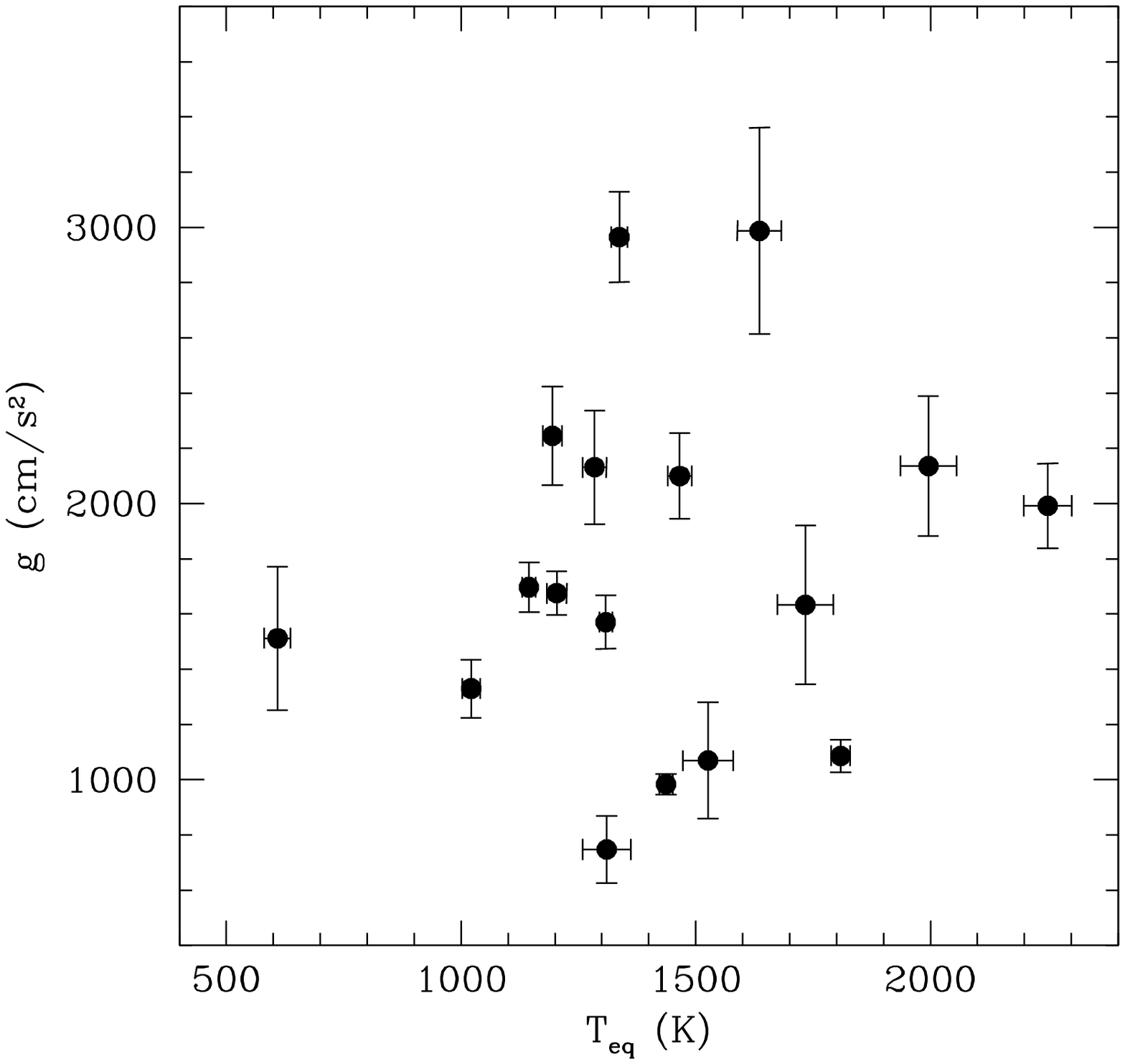}
\figcaption[f2.ps]{ The planet gravity plotted against equilibrium temperature, as defined in the text. Note that the
distribution is broader than when the gravity is plotted against period and that there appears to be a hint
of a gap between high and low gravity planets.  \label{gT}}
\newpage
\plotone{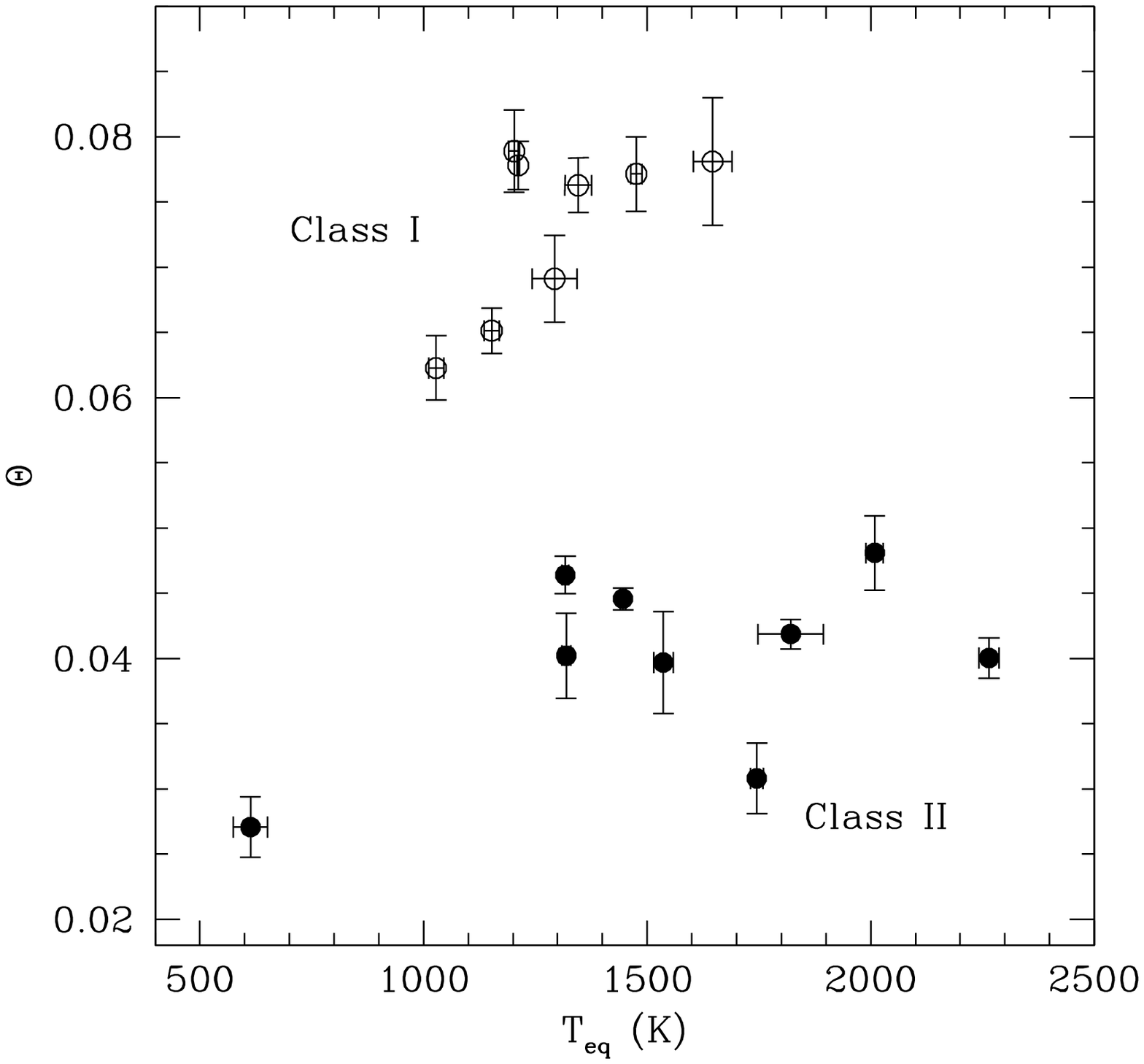}
\figcaption[f3.ps]{If we replace the gravity with the Safronov number $\Theta$, we find that there are now two
clear groups at fixed equilibrium temperature $T_{eq}$, apart from two outliers discussed in the text. We
label them as `Class~I' (open points) and `Class~II' (solid points).
 \label{OT}}
\newpage
\plotone{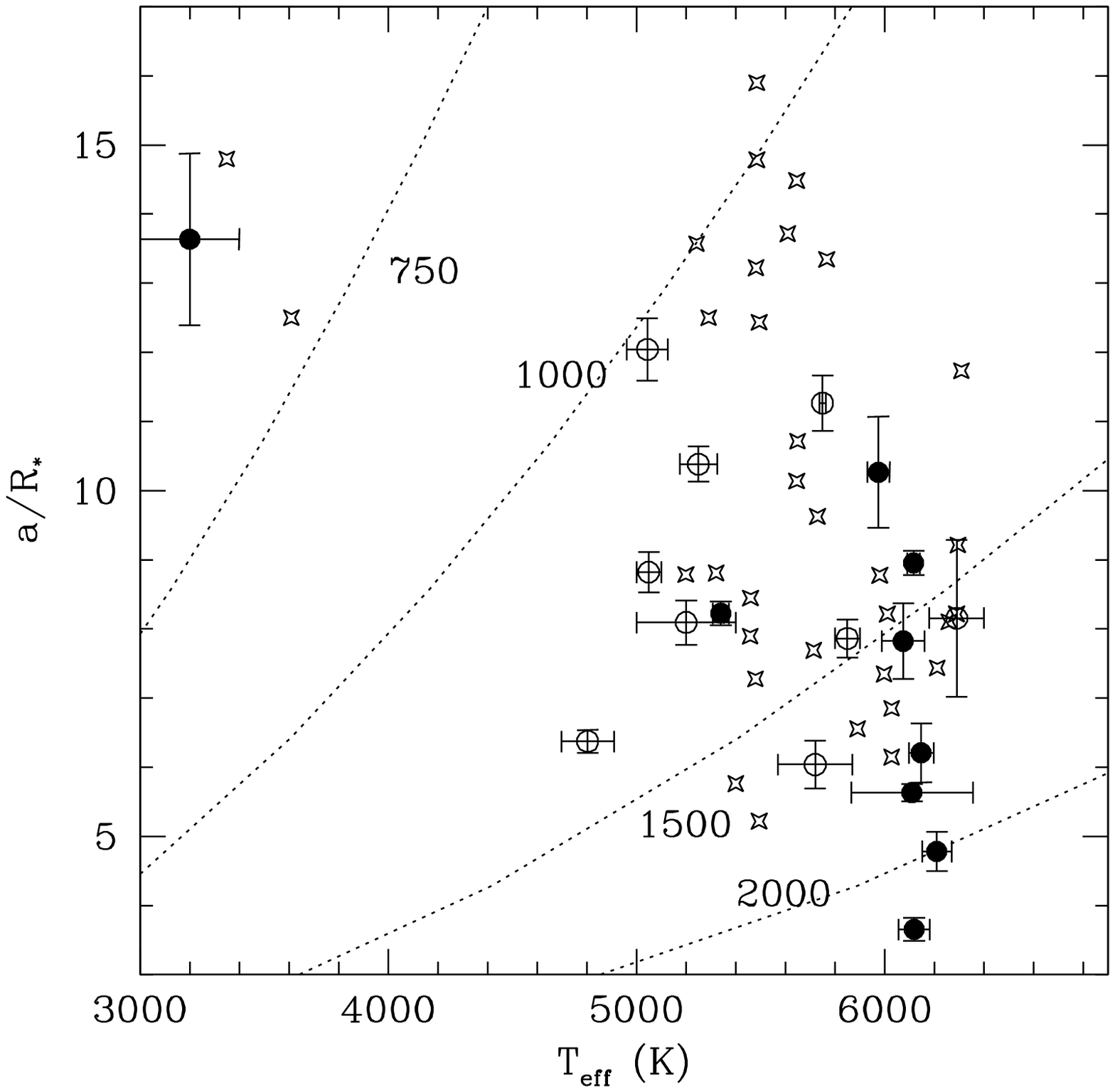}
\figcaption[f4.ps]{One might term this the `Illumination diagram', as the two quantities plotted (stellar effective
temperature and semi-major axis scaled by the stellar radius) determine how much stellar flux is received at the
surface of the planet.
Open and filled circles indicate Class~I and Class~II planets, as in Figure~\ref{OT}. To plot a planet on this diagram
one does not require the planetary mass or radius and so we can also plot the positions of all the other radial velocity planets 
as well. These are shown by the star symbols. Also shown are dotted lines indicating lines of constant $T_{eq}$.
The split into Class~I and Class~II is not as obvious in this diagram, although the general trend is for Class~II
to orbit hotter stars.
\label{Ta}}
\newpage
\plotone{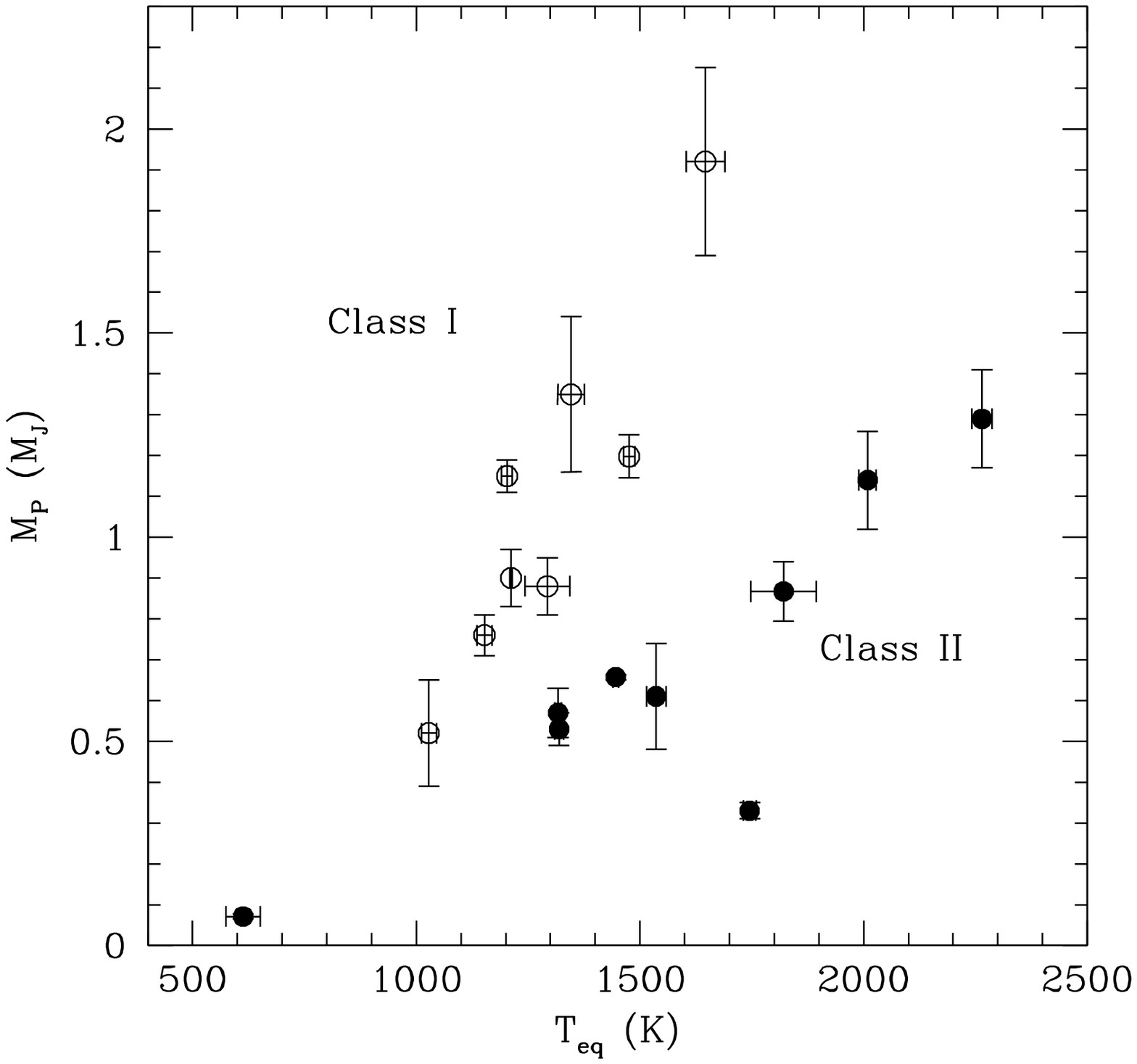}
\figcaption[f5.ps]{ Planet masses plotted against equilibrium temperatures. The two classes of planet
both appear to obey an approximately linear relation with $T_{eq}$, but with slopes different by almost
a factor of 3. \label{MT}}
\newpage
\plotone{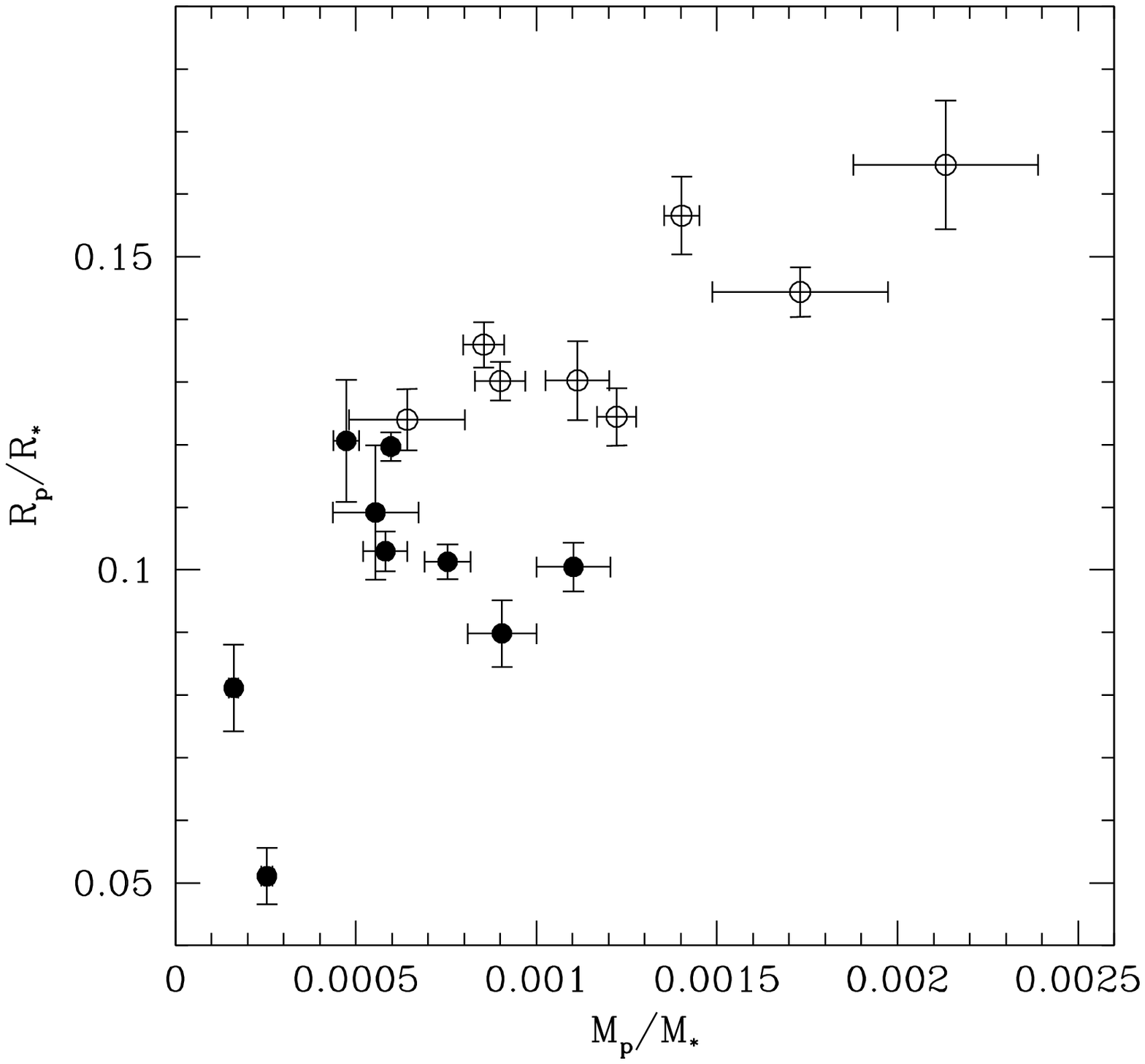}
\figcaption[f6.ps]{The open circles are Class~I and filled circles are Class~II. Rescaling an individual
star will not move a point in this diagram, suggesting that the observed dichotomy is real. The absence
of objects near ($9 \times 10^{-3}$,0.113) is curious, and unexplained. If this were indicative of some
systematic error in the measure of $R_p/R_*$ it would still not explain the dichotomy, since increasing
the radius of the Class~II planets would just make $\Theta$ smaller and similarly, the $\Theta$ of Class~I
would just get larger.
\label{RR}}
\newpage
\plotone{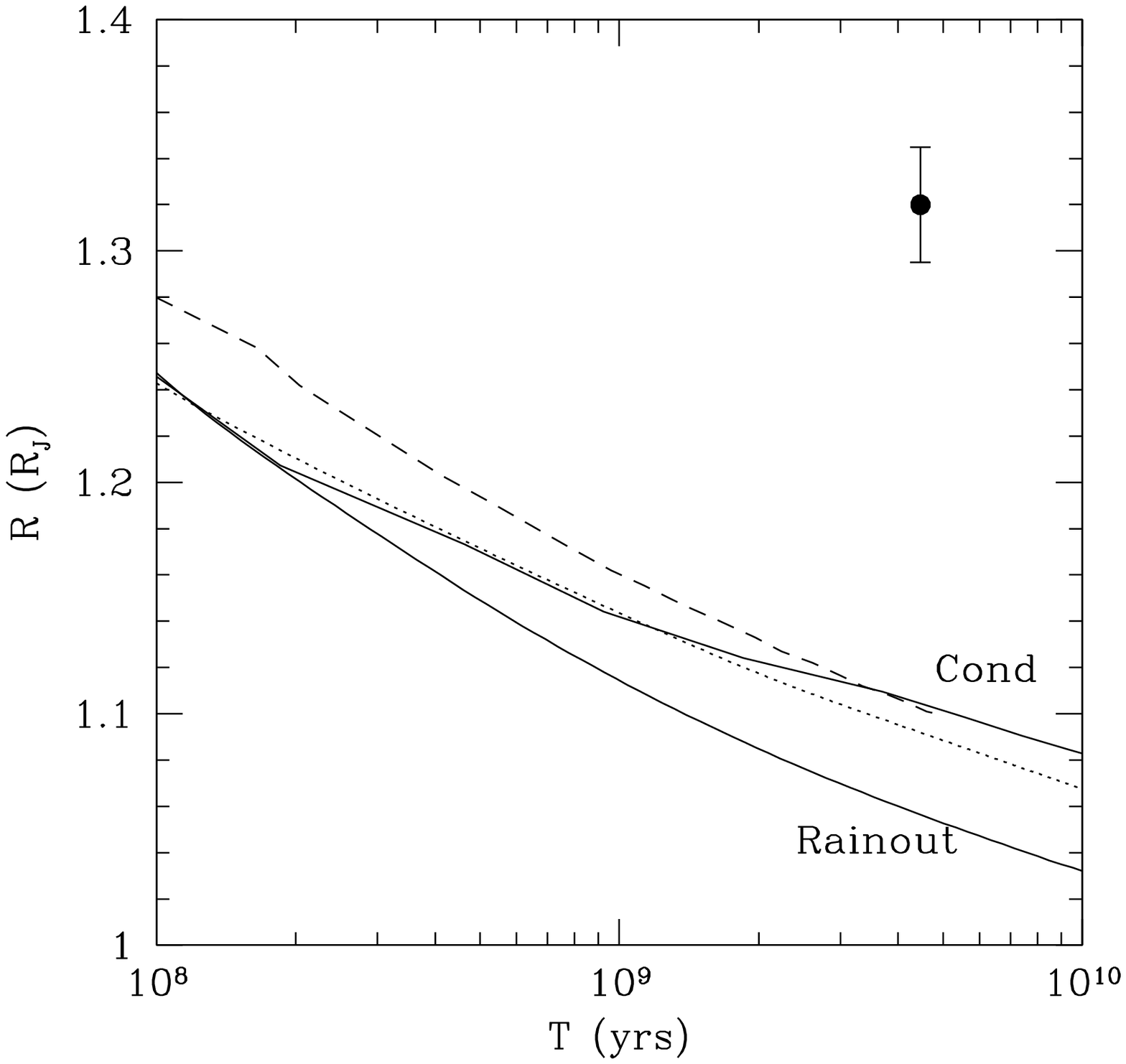}
\figcaption[f7.ps]{The dotted curve is the model of Baraffe et al. (2003), while the dashed curve is
from Burrows et al. (2007).
The upper solid curve is our model calculated using the same
boundary conditions as Baraffe et al. (termed the `cond' approximation).
 The lower solid curve is our model for the same planet, but with the boundary conditions calculated in the "rainout" (cloud-free) 
approximation, which we will use throughout the rest of the paper.
Planet mass is assumed to be $0.69 M_J$ in our models, as in the Baraffe model (the Burrows model is for a
0.64$M_J$ planet). The observed value of the planetary radius is shown in the upper right.
Note that we have not included the transit radius correction in this plot, since the principal goal is the
comparison of Henyey model results. \label{Verify}}
\newpage
\plotone{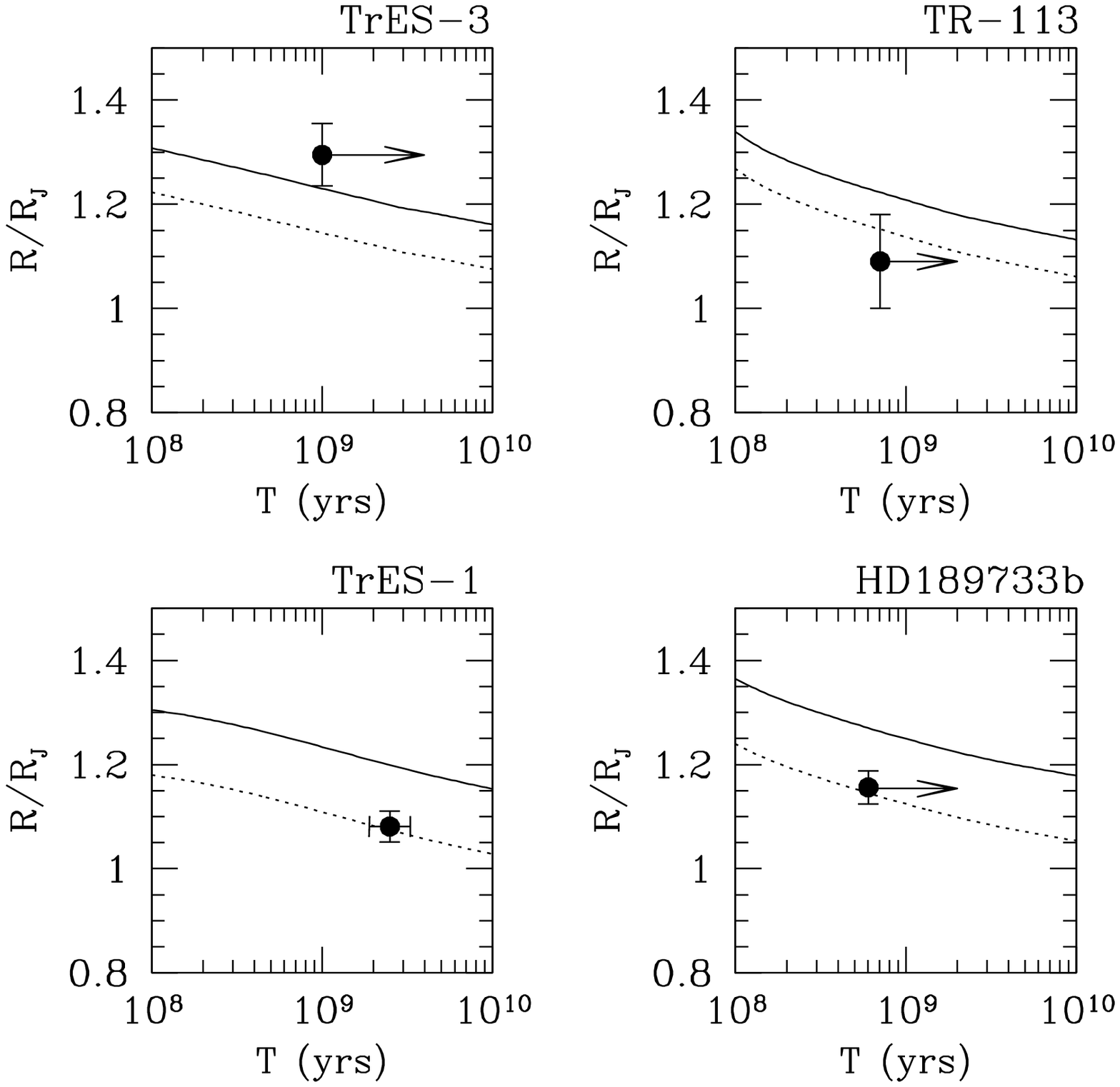}
\figcaption[f8.ps]{The four panels show evolutionary curves for four of the Class~I planets, using masses
and boundary conditions appropriate to each system. The dotted line indicates the traditional radius and the
solid curve indicates the expected optical transit radius. The measured planetary radii are also shown, and are
consistent with the models in all cases, as long as moderate heavy element cores are allowed.\label{Panel1}}
\newpage
\plotone{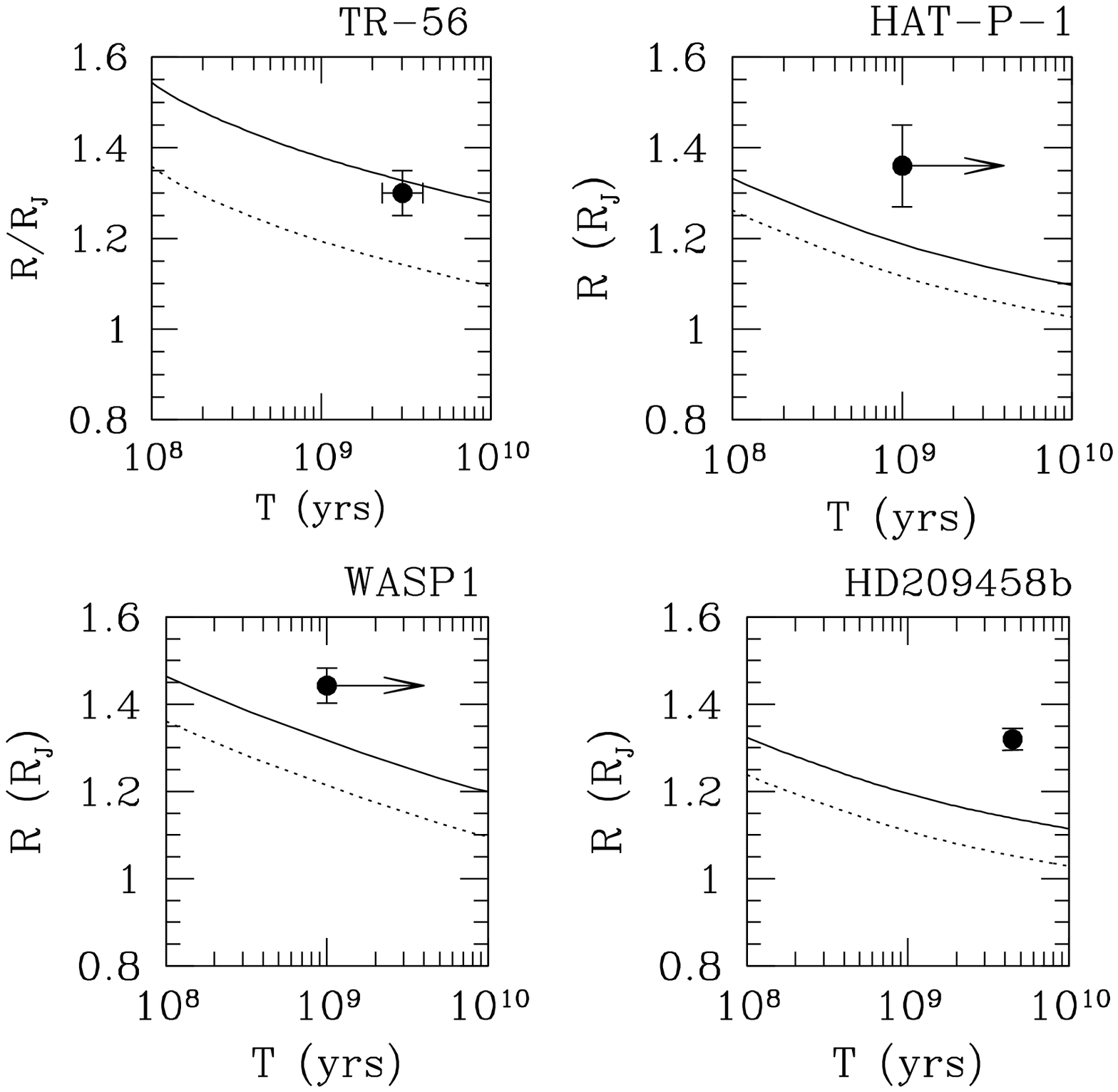}
\figcaption[f9.ps]{The four panels show evolutionary curves for four of the Class~II planets, using
masses and boundary conditions appropriate to each situation.  The dotted line indicates the traditional radius and the
solid curve indicates the expected optical transit radius. The measured planetary radii are also shown. 
Unlike the Class~I planets, several of these planets seem to be too large compared to the models, unless the planets are
implausibly young. Note also that the transit correction can vary from system to system, as it depends somewhat on the
level of illumination. \label{Panel2}}
\newpage
\plotone{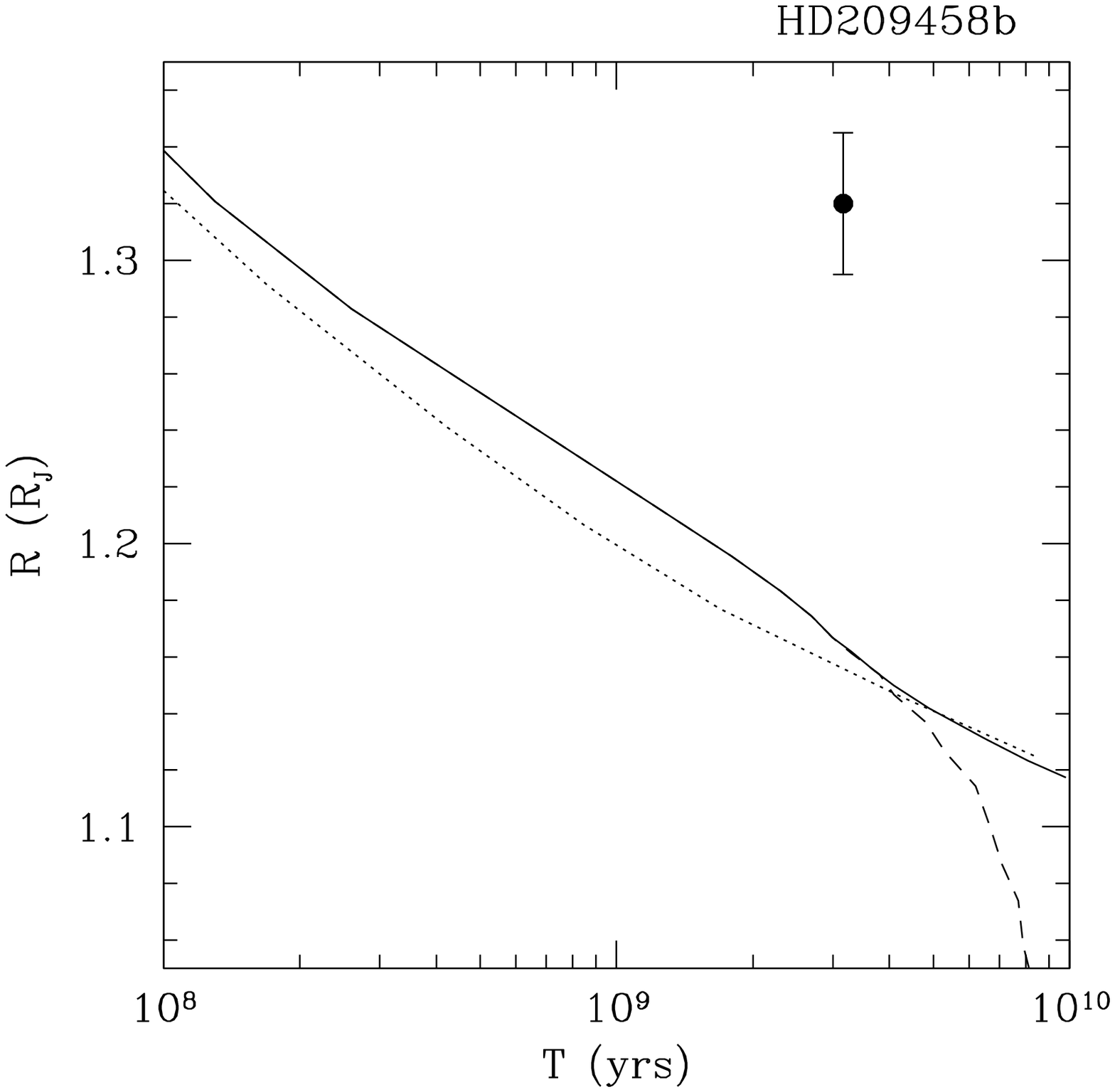}
\figcaption[f10.ps]{The solid curve shows the radius evolution of a planet evaporating from 1.1$M_J$ to 0.62$M_J$
over 3~Gyr. The dotted line indicates the cooling curve for a $0.62 M_J$ planet. The boundary conditions used are
appropriate for the HD209458b case and the filled circle indicates the observed value for that planet. \label{Evap}}
\newpage
\plotone{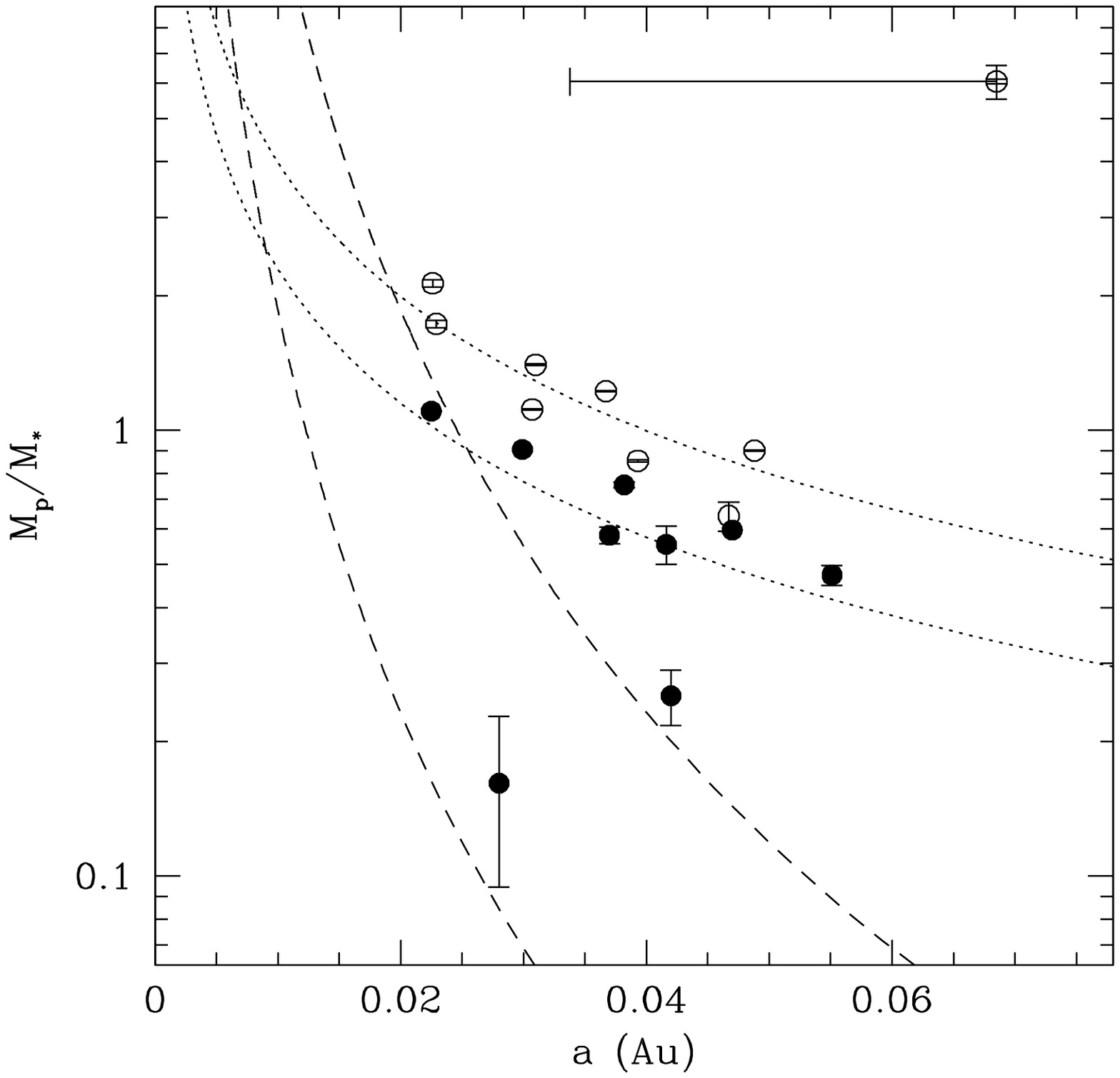}
\figcaption[f11.ps]{ Filled and open circles once again indicate Class~I and Class~II planets. For the eccentric transiting planet HD147506, we include an error bar to indicate the radial excursion between semi-major axis and
periastron.
 The dashed
lines indicate the Roche limit and twice the Roche limit. The dotted lines indicate lines of
constant Safronov number ($\Theta=0.04$ for the lower curve, $\Theta=0.07$ for the upper curve) assuming
a planetary radius of 1.2~$R_J$. \label{MMA}}
\newpage
\plotone{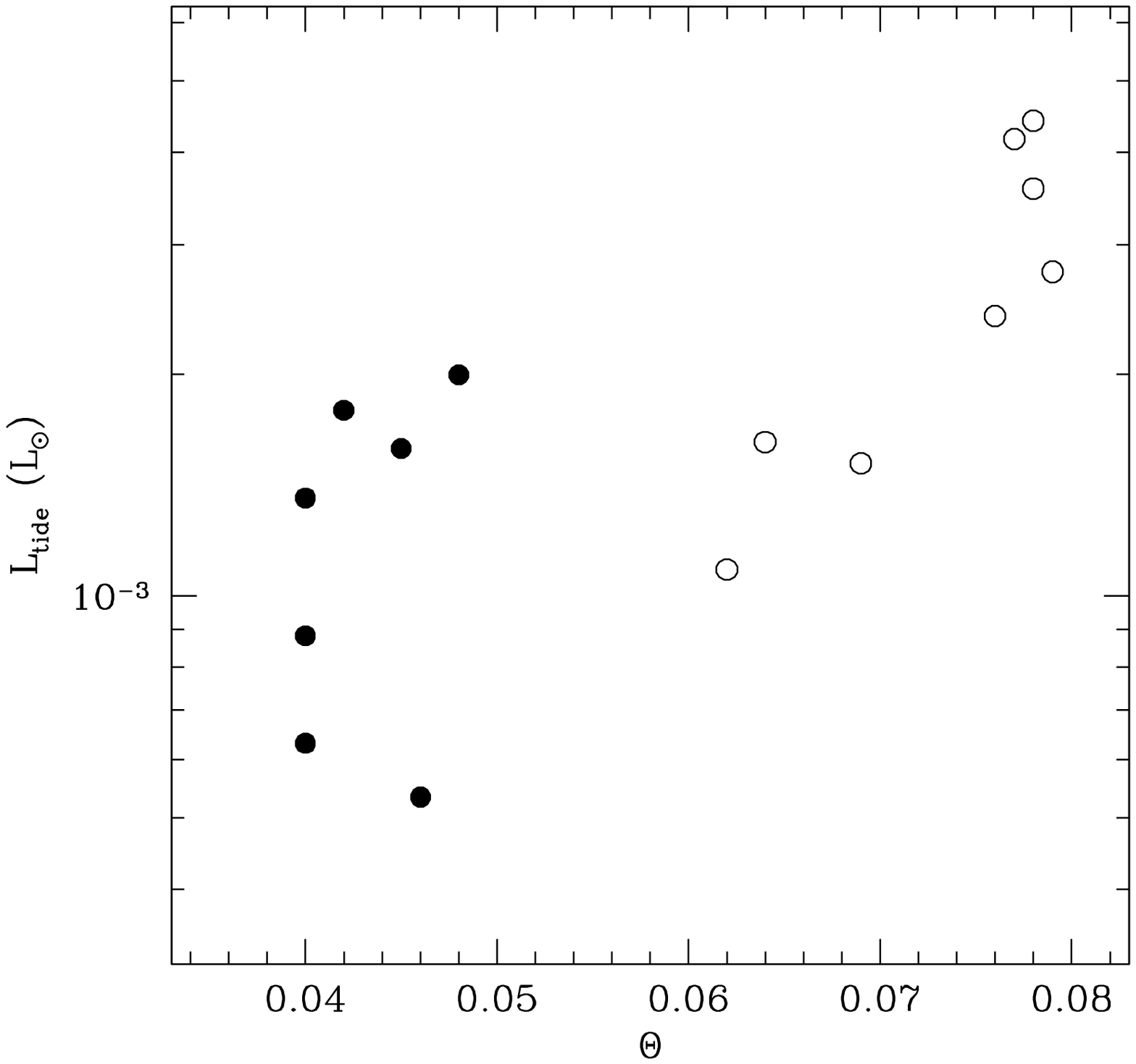}
\figcaption[f12.ps]{The open and solid points again indicate Class~I and Class~II planets respectively.
$L_{tide}$ is calculated as the internal luminosity required to inflate a H/He planet of the appropriate
mass to fill the Roche lobe of each planet at it's current location. (Note we have not plotted a point
for HD149026b, since that is clearly significantly different from a H/He planet.) \label{Tide}}
\newpage
\plotone{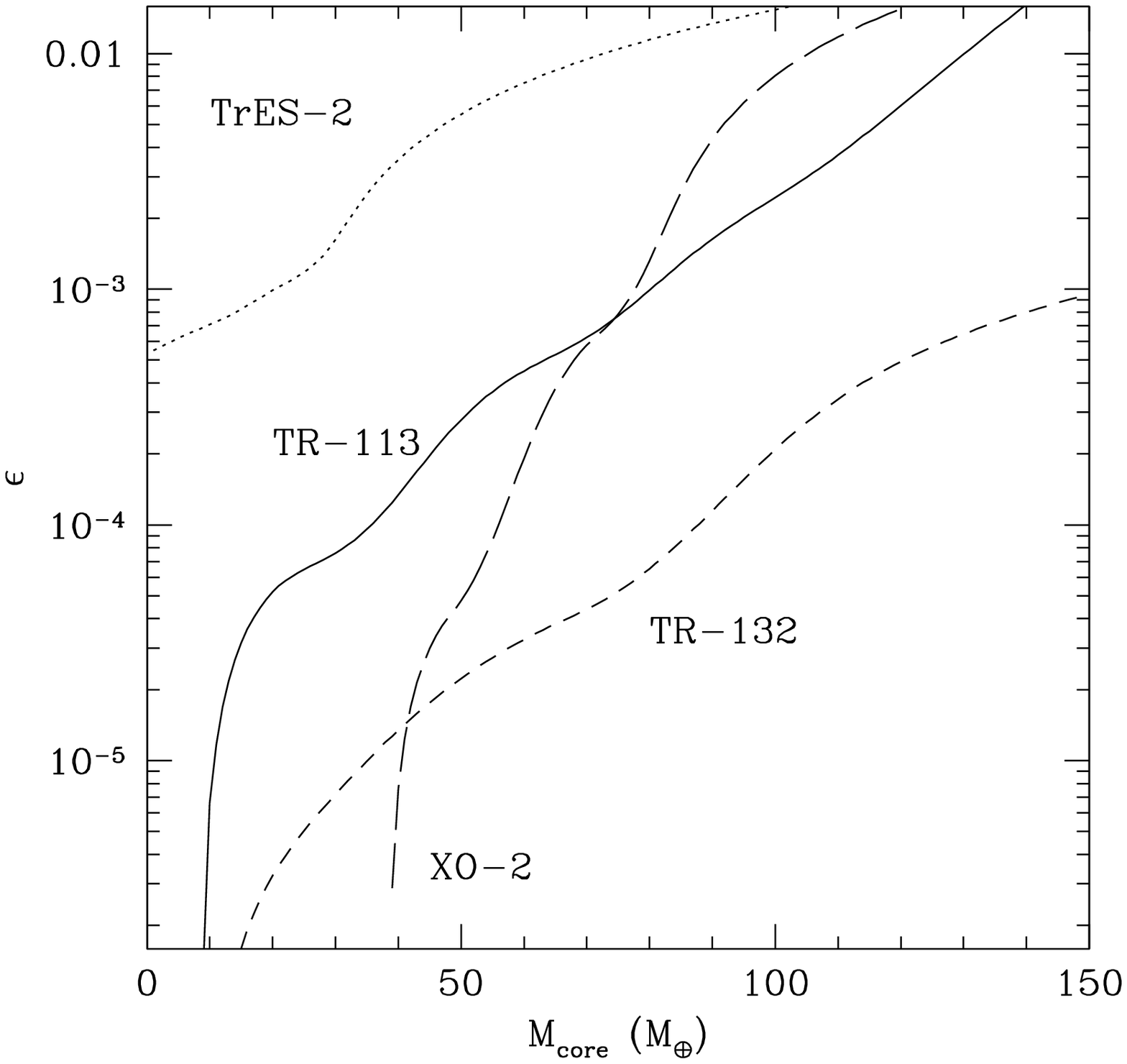}
\figcaption[f13.ps]{The four curves show the relation between insolation recycling efficiency $\epsilon$ and
core mass $M_{core}$, as determined for four different planetary systems. In each case, the values are
constrained so that the combination results in a model radius that matches the observed radius for that
system. 
 \label{EC}}
\newpage
\plotone{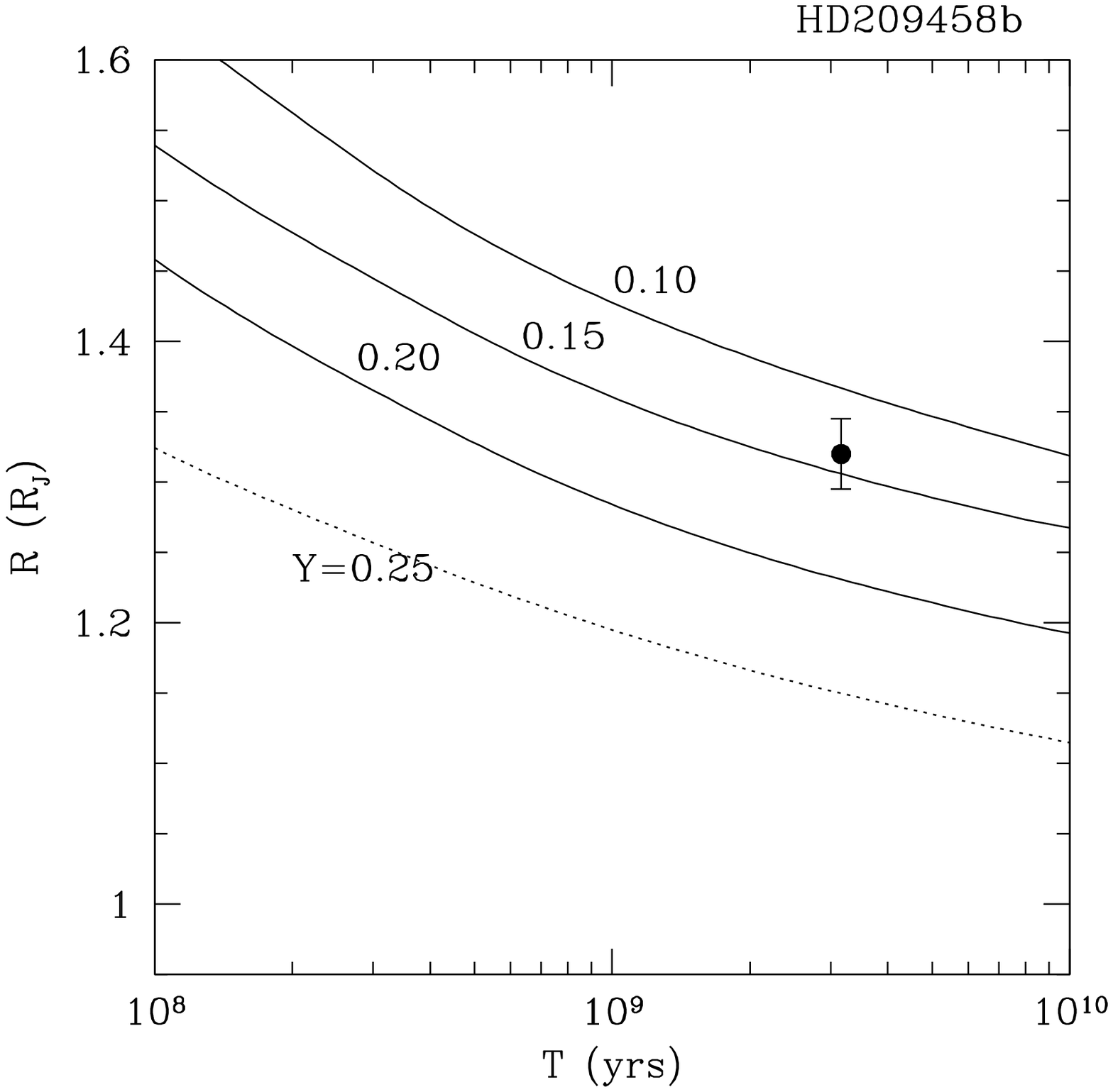}
\figcaption[f14.ps]{The dotted line is the evolution of a $0.65 M_J$ planet with a solar composition atmosphere
and envelope, irradiated in a manner appropriate to HD209458b. The solid lines indicate models in which a
fraction of the 
Helium has been removed from the envelope, leaving the planet with a Helium mass fraction Y as indicated.
\label{He209}}
\newpage
\plotone{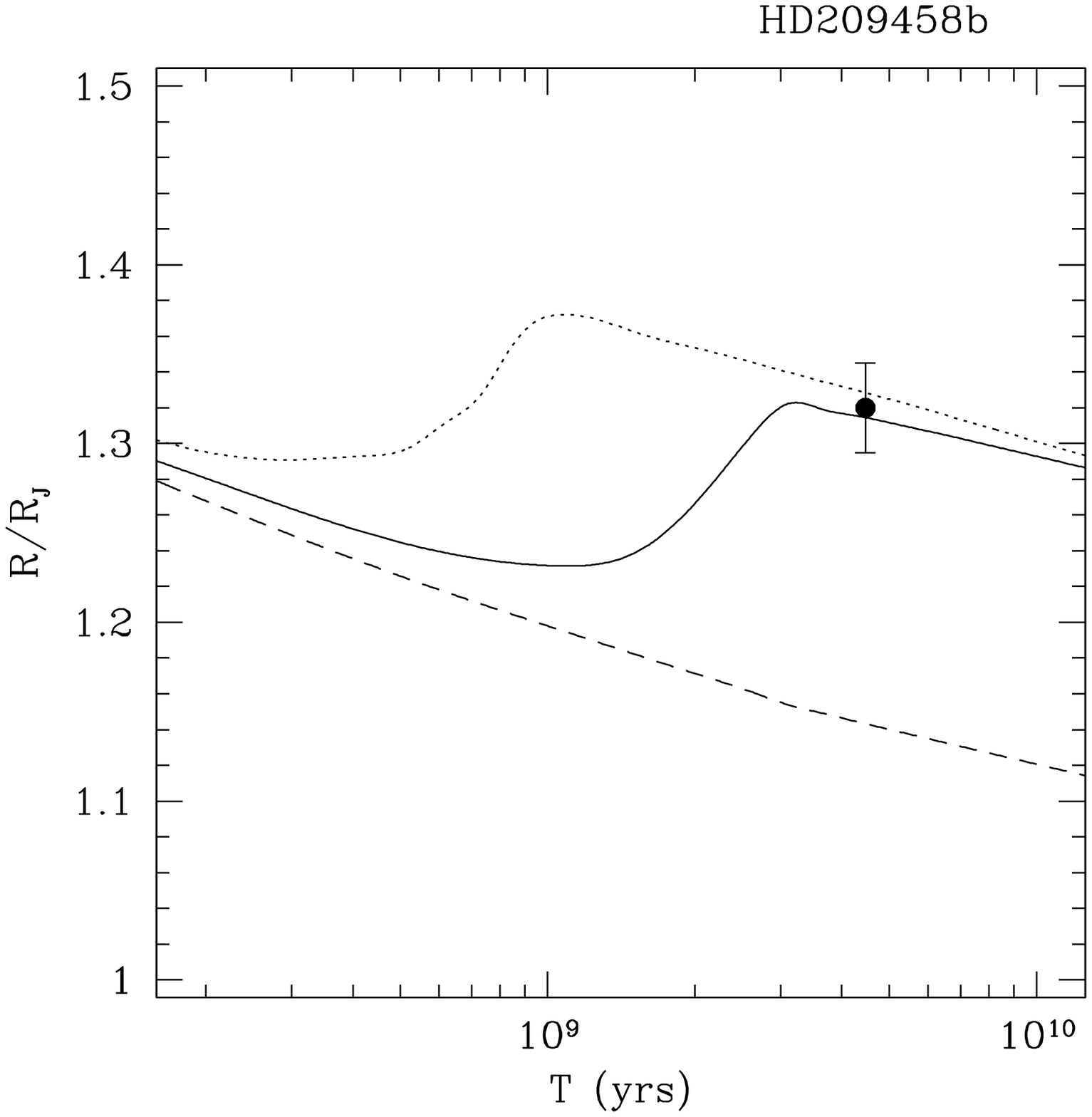}
\figcaption[f15.ps]{The solid line shows the evolution of a planet whose mass is reduced from 1~$M_J$
to 0.63~$M_J$ over the course of 3~Gyr. The mass removed is 50\% Helium by mass, so that the mean molecular
weight of the planet drops while mass is being lost. The dotted line shows the evolution if the same amount
of mass is lost over 1~Gyr. Finally, the dashed line shows the evolution if the mass loss occurs over 3~Gyr,
but the mass removed has the normal cosmic composition of Helium.
\label{EvapHe}}
\newpage
\plotone{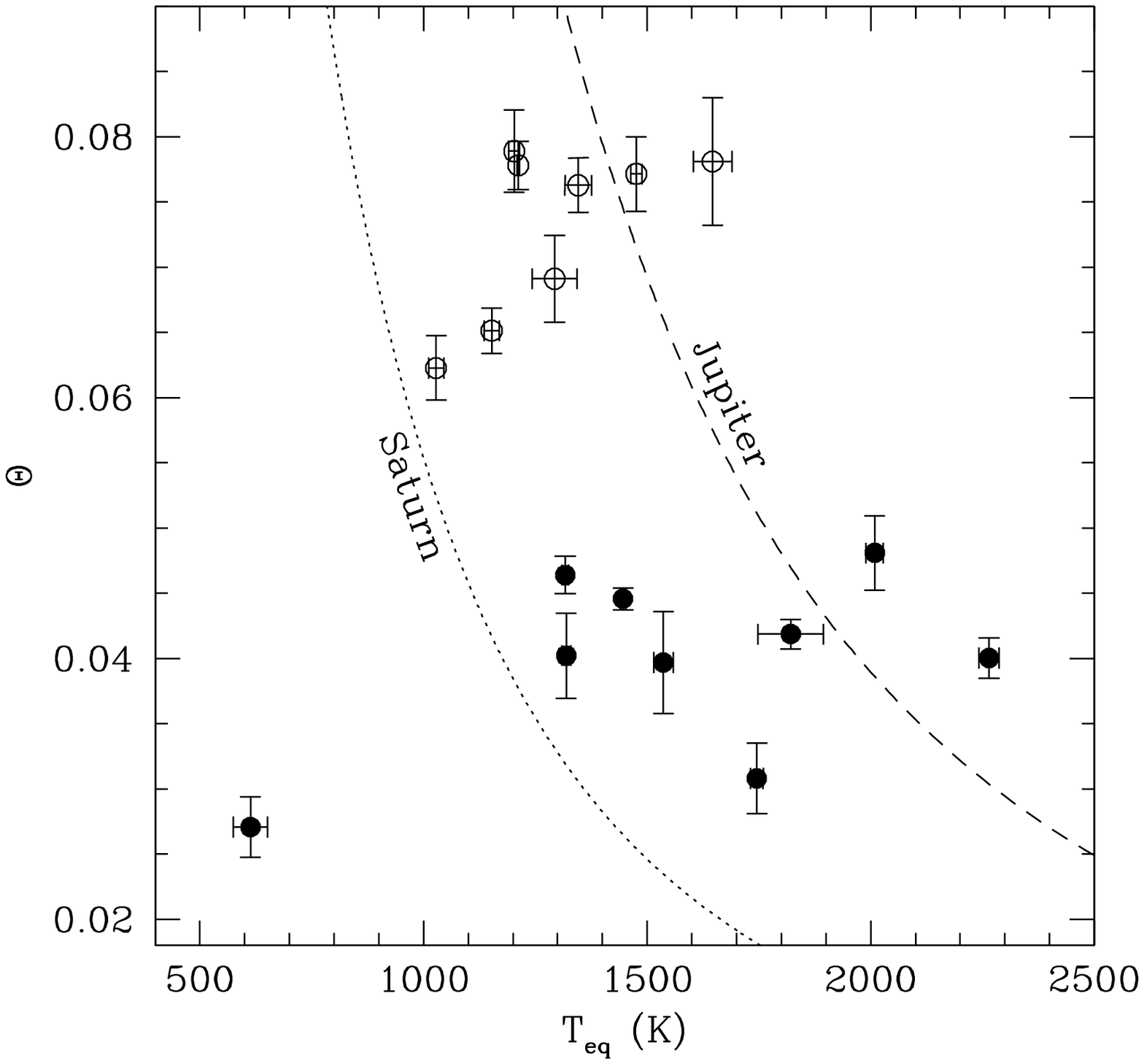}
\figcaption[f16.ps]{The dotted and dashed curves show how Jupiter and Saturn would move in the
$T_{eq}$--$\Theta$ diagram as one moved them closer to the sun. We have not adjusted the radii to
account for the insolation as it is a small effect on $\Theta$. \label{OTJS}}

\begin{references}
%\reference{BDB} Binney, J., Dehnen, W. \& Bertelli, G., 2000, MNRAS, 318, 658
\reference{B03} Baraffe, I., Chabrier, G., Barman, T., Allard, F., \& Hauschildt, P., 2003, A\&A, 402, 701 
\reference{B05} Baraffe, I., Chabrier, G., Barman, T., Selsis, F., Allard, F., \& Hauschildt, P., 2005, A\&A, 436, L47
\reference{B01} Barman, T. S., Hauschildt, P. H. \& Allard, F., 2001, ApJ, 556, 885
\reference{B02} Barman, T. S., Hauschildt, P. H.,  \& Allard, F., 2005, ApJ, 632, 1132
\reference{Bod} Bodenheimer, P., Lin, D. N. C. \& Mardling, R., 2001, ApJ, 548, 466
\reference{BI} Burkert, A.  \& Ida, S., 2007, ApJ, 660, 845
\reference{BHB} Burrows, A., Hubeny, I., Budaj, J. \& Hubbard, W. B., 2007, ApJ, 661, 502
\reference{BL93} Burrows, A. \& Liebert, J., 1993, Rev. Mod. Phys., 65, 301
\reference{BTR} Burrows, A., Sudarsky, D. \& Hubbard, W. B., 2003, ApJ, 594, 545
\reference{BMW} Butler, R. P., Marcy, G. W., Williams, E., Hauser, H. \& Shirts, P., 1997, ApJ, 474, L115
\reference{CB} Chabrier, G. \& Baraffe, I., 2007, ApJ, 661, L81
\reference{CBL} Charbonneau, D., Brown, T. M., Latham, D. W. \& Mayor, M., 2000, ApJ, 529, L45
\reference{DQT} Duncan, M. J., Quinn, T. \& Tremaine, S., 1987, AJ, 94, 1330
\reference{FI} Fernandez, J. A. \& Ip, W.-H., 1984, Icarus, 58, 109
\reference{FR96} Ford, E. B. \& Rasio, F. A., 1996, Science, 274, 954
\reference{FR} Ford, E. B. \& Rasio, F. A., 2006, ApJ, 638, L45
\reference{FH} Fortney, J. J. \& Hubbard, W. B., 2004, ApJ, 608, 1039
\reference{GaCOSPAR} Gautier, D., et al., 2006, Proc. 36th COSPAR General Assembly, 867
\reference{GW} Gessmann, C. K. \& Wood, B. J., 2002, Earth Planet. Sci. Lett., 200, 63
\reference{GP} Gillon, M., Pont, F., Demory, B.-O., Mallmann, F., Mayor, M., Mazeh, T., Queloz, D., Shporer, A.,
Udry, S. \& Vuissoz, C., 2007, arXiv0705.2219
\reference{GLB} Gu, P.-G., Lin, D. N. C. \& Bodenheimer, P., 2003, ApJ, 588, 509
\reference{GZ} Guillot, T., Santos, N. C., Pont, F., Iro, N., Melo, C. \& Ribas, I., 2006, A\&A, 453, L21
\reference{GS} Guillot, T. \& Showman, A. P., 2002, A\&A, 385, 156
\reference{H00} Hansen, B., 2000, arXiv:astro-ph/0004058
\reference{H96} Hansen, B., 1996, PhD Thesis, California Institute of Technology
\reference{H99} Hansen, B., 1999, ApJ, 580, 620
\reference{H06} Harrington, J., Hansen, B., Luszcz, S., Seager, S., Deming, D., Menou, K., Cho, J.
\& Richardson, L. J., 2006, Science, 314, 623
\reference{H07} Harrington, J., Luszcz, S., Seager, S., Deming, D. \& Richardson, L. J.,  2007, Nature, 447, 691
\reference{EB} Hauschildt, P. \& Baron, E., 1999, J. Comp. Appl. Math., 109, 41
\reference{HM} Henry, G. W., Marcy, G. W., Butler, R. P. \& Vogt, S. S., 2000, ApJ, 529, L41
\reference{KL} Khodachenko, M. L., et al., 2007, Planet. Space Sci., 55, 631
\reference{KCA} Knutson, H. A., Charbonneau, D., Allen, L. E., et al., 2007, Nature, recently
\reference{LCOROT} Lammer, H. , et al., 2007, arXiV:astro-ph/0701565
\reference{LBM} Lin, D. N. C., Bodenheimer, P. \& Mardling, R., 2001, 548, 466
\reference{LBR} Lin, D. N. C., Bodenheimer, P. \& Richardson, D. C., 1996, Nature, 380, 606 
\reference{MQ} Mayor, M. \& Queloz, D., 1995, Nature, 378, 355
\reference{MZP} Mazeh, T., Zucker, S. \& Pont, F., 2005, MNRAS, 356, 955
\reference{MSP} Melo, C., Santos, N. C., Pont, F., Guillot, T., Israelian, G., Mayor, M., Queloz, D. \&
Udry, S., 2006, A\&A, 460, 251
\reference{MHHT} Murray, N., Hansen, B., Holman, M. \& Tremaine, S., 1998, Science, 279, 69
%\reference{MH} Murtagh, F. \& Heck, A., 1986, Multivariate Data Analysis (Kluwer Academic)
\reference{PGMG} Pollack, J. B., Grossman, A. S., Moore, R. \& Graboske, H. V., 1977, Icarus, 30, 111
\reference{S72} Safronov, V. S., 1972, Evolution of the Protoplanetary Cloud and Formation of the
Earth and Planets (Israel Program for Scientific Translation, Jerusalem)
\reference{SCVH} Saumon, D., Chabrier, G. \& Van Horn, H. M., 1995, ApJS, 99, 713
\reference{SG} Showman, A. \& Guillot, T., 2002, A\&A, 385, 166
\reference{SWS} Southworth, J., Wheatley, P. J. \& Sams, G., 2007, MNRAS, arXiV:0704.1570
\reference{SS} Stevenson, D. J. \& Salpeter, E. E., 1977, ApJS, 35, 221
\reference{T93} Tremaine, S., 1993, in Planets Around Pulsars, ed. J. A. Phillips, S. E. Thorsett \& S. R. Kulkarni, 335
\reference{Trill} Trilling, D. E., Benz, W., Guillot, T., Lunine, J. I., Hubbard, W. B. \& Burrows, A., 1998,
ApJ, 500, 428
\reference{VM} Vidal-Madjar, A., Lecavelier des Etangs, A., Desert, J. -M., Ballester, G. E., Ferlet, R., Hebrand, G. \& Mayor, M., 2003, Nature, 422, 143
\reference{WH} Winn, J. \& Holman, M. J., 2005, ApJ, 628, L159
\reference{Yelle} Yelle, R. V., 2004, Icarus, 170, 167
\end{references}
\end{document}